\documentclass[journal]{IEEEtran}
\IEEEoverridecommandlockouts
\usepackage{tikz}
\usepackage{lipsum}
\usepackage{graphicx}
\usepackage{microtype}
\usepackage{multicol}
\usepackage{subcaption}
\usepackage{makecell}  % in preamble

\usepackage{enumitem} 
\usetikzlibrary{decorations.pathreplacing, positioning}

\usetikzlibrary{shapes.geometric, arrows}

\tikzstyle{startstop} = [rectangle, rounded corners, minimum width=2cm, minimum height=1cm, text centered, draw=black, fill=red!30]
\tikzstyle{process} = [rectangle, minimum width=2cm, minimum height=1cm, text centered, draw=black, fill=white, align=center]
\tikzstyle{arrow} = [thick,->,>=stealth]

\usepackage{ifpdf}
\usepackage{cite}
\usepackage{graphicx}
\usepackage{amsmath}
\usepackage{algorithmic}
\usepackage{array}
\usepackage{cite}
\usepackage{amsmath,amssymb,amsfonts}
\usepackage{hyperref}
\usepackage{algorithmic}
\usepackage{graphicx}
\usepackage{textcomp}
\usepackage{xcolor}
\usepackage{tikz}
\usepackage{multirow} % For merging cells

\usetikzlibrary{matrix, shapes.geometric, arrows}

\ifCLASSOPTIONcompsoc
  \usepackage[caption=false,font=normalsize,labelfont=sf,textfont=sf]{subfig}
\else
  \usepackage[caption=false,font=footnotesize]{subfig}
\fi
\usepackage{fixltx2e}
 \usepackage{dblfloatfix}
\usepackage{url}
\usepackage{orcidlink}  % Add this package

\usepackage{tikz}
\usetikzlibrary{shapes.geometric, arrows.meta, positioning}

\usepackage[final]{changes}

\definecolor{clara}{rgb}{1,0,1}  % Magenta

\usepackage[normalem]{ulem}
\makeatletter
\providecommand{\sf@counterlist}{}

\let\@LN@col\relax
\makeatother

\usepackage[switch]{lineno}

\makeatletter
\def\ps@copyrightpage{%
  \def\@oddhead{%
    \vbox{%
      \hbox to \textwidth{%
        \parbox[t]{\textwidth}{%
          \centering
          \scriptsize
          \copyright~2026 IEEE. Personal use of this material is permitted.  Permission from IEEE must be obtained for all other uses, in any current or future media, including reprinting/republishing this material for advertising or promotional purposes, creating new collective works, for resale or redistribution to servers or lists, or reuse of any copyrighted component of this work in other works.
        }%
        \hfil
      }%
      \vskip 8pt
    }%
  }%
  \def\@evenhead{\@oddhead}%
  \def\@oddfoot{\hfil\thepage\hfil}%
  \def\@evenfoot{\hfil\thepage\hfil}%
}
\makeatother

\begin{document}
% \linenumbers
%\title{A Ridge-Guided Deep Learning Denoiser for Enhancing Ultrasonic Vocalization Analysis}

%\title{Vocalization Denoising Using Synthetic Clean Signals}

\title{Training Set Synthesis for Bioacoustic Denoising: \\ A Case Study with Mice}


\author{Reyhaneh~Abbasi\,\orcidlink{0000-0002-8423-6672},
        Peter~Balazs\,\orcidlink{0000-0002-8423-6672},
        Vincent~Lostanlen\,\orcidlink{0000-0002-8423-6672},
        ~Clara~Hollomey\,\orcidlink{0000-0002-8423-6672},
        ~Dustin~J. Penn\,\orcidlink{0000-0002-8423-6672},
        Sarah M.~Zala\,\orcidlink{0000-0002-0698-5299},
        Nicki~Holighaus\,\orcidlink{0000-0002-8423-6672}
        % <-this % stops a space
\thanks{%Part of this
%work was presented at the IEEE International Conference on Acoustics, Speech, and Signal Processing 2025 [1].
(Corresponding author: Reyhaneh Abbasi, reyhaneh.abbasi@oeaw.ac.at).
Reyhaneh Abbasi, Peter Balazs, and Nicki Holighaus are at the Acoustics Research Institute of the Austrian Academy of Sciences,
Vienna, Austria. Peter Balazs is also affiliated with the Interdisciplinary Transformation University Austria (IT:U), Linz, Austria.
Reyhaneh Abbasi is also affiliated with the Vienna Doctoral School of Cognition, Behavior, and Neuroscience, University of Vienna, Vienna, Austria.
Vincent Lostanlen is at the
Nantes Université, École centrale Nantes
CNRS, LS2N, UMR 6004
F-44000 Nantes, France.
Clara Hollomey is at the Institute for Creative Media Technologies of the University of Applied Sciences Saint Pölten. Dustin J. Penn and Sarah M. Zala are at the Konrad Lorenz Institute of Ethology, Department of Interdisciplinary Life Sciences, University of Veterinary Medicine, Vienna, Austria.}% <-this % stops a space
}

\maketitle
\thispagestyle{copyrightpage}

\begin{abstract}
Bioacoustic recordings are often degraded by ambient noise, which complicates the analysis of weak or noise-overlapped vocalizations. Convolutional neural networks, particularly U-Net architectures, have shown a strong denoising performance in speech and music processing. However, their direct application to bioacoustic signals is limited by the scarcity of clean training data. To address this issue, we propose a training set synthesis approach and develop a supervised denoising model that predicts a complex ratio mask in the time-frequency domain. The model leverages ridges, or frequency contours, that represent the fundamental frequency together with one or more harmonic partial components of vocalizations. These ridges are used both for the synthesis of training sets and to design a loss function that assigns higher weights to the ridge regions (ridge-guided loss function). This weighting step helps the network better preserve vocalization details during denoising. As a case study, we evaluate our approach using ultrasonic vocalizations (USVs) recordings of house mice, which are widely studied in behavioral biology and neuroscience. In actual field recordings, the proposed method enhances fundamental and harmonic partial ridge tracking compared to our previous signal-processing approach. In addition, a classifier trained on denoised data improves USV classification on out-of-sample, noisy recordings from wild and domesticated mice compared to classifiers trained on noisy recordings. Our proposed method also substantially improves the scale-invariant signal-to-distortion ratio on synthetic testing data across a wide range of input signal-to-noise ratios. Although we focus on USVs, the proposed approach should be broadly applicable to other bioacoustic signals with trackable ridges, and thus enables ridge-based training set synthesis and denoising.
\end{abstract}


% \fi
\IEEEpeerreviewmaketitle

%%%%%%%%%%%%%%%%%%%
%Introduction
%%%%%%%%%%%%%%%%%%
\section{Introduction}
Many animals, including birds, bats, and rodents, rely on vocalizations for communication \cite{marler2004nature, schnitzler2001echolocation, portfors2007types}. However, acoustic signals are often degraded by ambient noise, which complicates their analysis and interpretation \cite{bradbury1998principles}. Traditional denoising techniques, such as spectral subtraction~\cite{boll2003suppression}, estimate the noise spectrum during vocal pauses and subtract it from the rest of the signal. These methods work well for simple, consistent noise, such as white noise, but they assume that the noise remains relatively stable over time and across frequencies. However, in complex natural environments, where noise levels and their spectral patterns can vary rapidly, these assumptions may not hold~\cite{xie2021bioacoustic}. 

Recent progress in machine learning provides new approaches to denoising bioacoustic recordings \cite{yang2021transfer, vickers2021robust}. Supervised models, which rely on paired noisy-clean audio for training, have shown strong performance in related domains. In bioacoustics, however, large datasets of clean recordings are rarely available, which limits the applicability of such methods. A common machine-learning approach to address this problem is to synthesize a training set by generating artificial or augmented clean signals (e.g., in environmental sound analysis)~\cite{ronchini2024synthetic,weldy2025simulated}. In the present study, we develop an approach for synthesizing paired training data for supervised denoising.

Acoustic signals, whether they are human speech, vocalizations of other animals, or others, are often analyzed using a time-frequency representation (e.g., spectrogram) to visualize and analyze temporal changes in frequency components. Tracking ridges — also called frequency tracks or contours — is an important task for extracting features from these time-frequency representations. Formally, at time \(n\), a ridge corresponds to a local (or global) maximum of the magnitude (or power) of the time-frequency representation with respect to the frequency axis \(k\). 

Depending on the signal, ridges may correspond to the fundamental frequency alone, the fundamental together with one or more harmonic partials, or other prominent tonal structures. Ridges have been widely used in bioacoustics for tasks such as age estimation \cite{stoeger2014age}, syllable-type classification \cite{zala2020primed}, and individual identification \cite{marconi2020ultrasonic,carlson2020individual,ceugniet2004vocal}.  

Many animal vocalizations can be modeled by locally sinusoidal components whose instantaneous frequency and amplitude tracks capture relevant signal information. Motivated by this structure, we propose a novel solution for denoising bioacoustic recordings, which uses automatically tracked ridge information both to synthesize a training set and to integrate a ridge-guided loss into the denoiser. Ridges are tracked using multi-component tracking \cite{abbasi2025robust}. 
%and synthesizes clean training data to guide the denoising process. 
Our approach provides two main contributions:
\begin{description}
\item[Training set synthesis.] We use automatically tracked ridge information (based on \cite{abbasi2025robust}) to synthesize a training set and train the denoiser. Importantly, the tracked ridges do not require manual correction and only need to be moderately accurate to be useful for the synthesis of the training set. 
\item[Ridge-guided loss.] We also use the ridges to weight the loss function, which helps to preserve the vocalization details while suppressing noise. 
\end{description}

This work presents the first application of ridge-based approaches to the synthesis of training sets and the weighting of loss functions for bioacoustic analysis. As a case study, we use the recordings of ultrasonic vocalizations (USVs) of male house mice (\textit{Mus musculus musculus}) \cite{klaus2025courtship}. Our approach is expected to be applicable to bioacoustic signals that are non-stationary and locally sinusoidal. These signals must have trackable ridges in their time-frequency representations, and the ridges should be representative of the vocalizations.
This includes tonal vocalizations such as mouse USVs and similar signals observed in other species, including marmoset vocalizations \cite{marm_audio_dataset} and dolphin whistles \cite{best_dataset}. In contrast, broadband or strongly noise-like vocalizations, for which ridge structure is not well defined, fall outside the scope of the proposed method.

In the context of mouse USVs, background noise can affect nearly every stage of analysis. For example, high-signal-to-noise ratio (SNR) signals are required for playback experiments testing the influence of USVs in mice~\cite{asaba2017male}. Similarly, accurate ridge-tracking is essential for computing and analyzing spectro-temporal features (e.g., duration, frequency modulation, frequency slope, and amplitude) \cite{nicolakis2020ultrasonic,chabout2015male}, because it isolates the dominant time-frequency trajectories of each vocalization. In addition, reliable classification depends on the quality of the USV signals. Consequently, noise complicates signal processing tasks and affects downstream analyses.

This work represents the first application of a deep learning-based denoiser to mouse USVs and the first systematic evaluation of its effects on USV ridge tracking and classification.
Since actual field recordings lack clean references, we synthesize a test set to compute scale-invariant signal-to-distortion ratio (SI-SDR) improvements. 
Additionally, improvements in ridge-tracking and classification accuracy of actual field recordings are used as task-based performance metrics to assess the effectiveness of our denoising approach. Ridge-tracking is evaluated against post hoc manual ridge annotations, while classification performance is assessed against manual USV labels. The proposed denoiser effectively improves SI-SDR for both high- and low-SNR synthetic noisy recordings relative to the corresponding synthetic clean signals. It also enhances the precision and recall of ridge-tracking, while reduces frequency deviations relative to manual ridge annotations, outperforming previous methods. The integrated denoising-and-retraining pipeline, which uses the proposed denoiser to preprocess audio for classifier retraining, generalizes to previously unseen data and improves classification F1-scores on both in-sample and out-of-sample noisy data. Overall, our proposed method substantially improves the reliability and accuracy of USV analysis in noisy environments without requiring clean or hand-annotated training data.

The next section (Section \ref{sec:relatedwork}) reviews related work on denoising approaches for bioacoustic signals in general and USVs in particular. Section \ref{sec:method} describes the proposed approach for the training set synthesis and denoising, and Section \ref{sec:experiment} outlines the experimental setup. Section \ref{sec:results} presents the results and discussion, and Section \ref{sec:discussion} concludes with potential future directions.

%%%%%%%%%%%%%%%%%
%Related Work
%%%%%%%%%%%%%%%%%%
\section{Related Work} \label{sec:relatedwork}
Understanding animal communication and behavior is the main aim of studies in bioacoustics. While the detection and analysis of animal vocalizations is relatively straightforward for recordings made in low-noise (high-SNR) conditions, noisy recordings present significant technical challenges \cite{xie2021bioacoustic}. In such cases, robust and effective denoising techniques are essential for reliable analyses. Despite its importance, denoising in bioacoustic applications remains relatively underexplored \cite{xie2021bioacoustic}. Below, we briefly review general bioacoustic denoising methods and categorize them into conventional (Section \ref{sec:relwork_conv}) and deep learning-based (Section \ref{sec:relwork_deep}) approaches. We also include ridge-tracking approaches, as our training set synthesis relies on ridge-tracking (Section \ref{sec:relwork_ridge}).

%%%%%%%%%%%%%%%%%
%Conventional Denoising Methods in Bioacoustics
%%%%%%%%%%%%%%%%%
\subsection{Conventional Denoising Methods in Bioacoustics}\label{sec:relwork_conv}
Denoising bioacoustic signals classically begins by selecting the relevant frequency range, followed by applying traditional signal-processing techniques. Popular methods include spectral subtraction \cite{boll2003suppression}, Wiener filtering \cite{scalart1996speech}, wavelet-transform-based techniques \cite{priyadarshani2016birdsong}, image-processing-based noise reduction \cite{zeppelzauer2013acoustic}, and more recent spectral-gating implementations such as noisereduce \cite{sainburg2024noisereduce}, which are widely used due to their simplicity and computational efficiency. These methods do not require clean training data for development, though such data may be needed in practice to tune hyperparameters or thresholds. Moreover, these methods rely on assumptions about noise characteristics, such as stationarity (e.g., white noise throughout a recording), which often are not met in natural acoustic environments. As a result, such methods may perform poorly and inadvertently remove subtle but important bioacoustic features \cite{xie2021bioacoustic}.

Processing mouse USVs is challenging because ambient noise interferes with call detection, acoustic-feature quantification, and syllable classification. Part of the noise has approximately equal intensity across all frequencies (i.e., white noise) and can be reduced using spectral subtraction \cite{van2017mupet}. The rest of the noise is generated by the mice, especially during locomotion and social interactions, when they primarily vocalize. Some of this noise lies below the USV frequency range and can be effectively reduced using band-pass filtering (e.g., 40-120 kHz) to isolate the relevant signals. However, noise components that overlap with the USV frequency range degrade the quality of signal detection and analysis, and similar issues also arise when the vocalizations themselves are weak or faint, i.e., near the noise floor.
 
To address these challenges, \cite{tachibana2020usvseg} developed a method that combines multitaper spectrograms with spectrogram flattening. Recent work \cite{abbasi2025robust} combines empirical Wiener shrinkage \cite{ghael1997improved}, multitapering \cite{xiao2007multitaper}, and time--frequency reassignment \cite{aufl95} to enhance the signal. Despite these advances, overlapping noise and faint USVs remain challenging and often degrade subsequent USV analyses \cite{abbasi2025bioacoustic}.

%%%%%%%%%%%%%%%%
%Denoising Methods for Mouse USVs
%%%%%%%%%%%%%%%%
\subsection{Deep Learning Approaches for Bioacoustic Denoising}\label{sec:relwork_deep}

Deep learning has become increasingly popular for denoising audio and speech recordings due to its ability to learn complex, nonlinear relationships between noisy and clean training data \cite{bulut2020low}. Unlike classical denoising algorithms, which rely on predefined signal models and prior assumptions, deep learning-based methods learn noise removal patterns from large training datasets. These models can perform adaptive denoising without requiring precise modeling of the signal and noise components, or manual parameter tuning \cite{yu2019deep}. 

In bioacoustic signal processing, deep learning has been applied to tasks such as call detection \cite{stowell2019automatic,goussha2022hybridmouse}, syllable-type classification \cite{schwab2023automated}, and species classification \cite{lostanlen2024birdvoxdetect}. However, applications of deep learning to denoising bioacoustic signals remain relatively limited \cite{zhang2024automatic}, with only a few studies exploring this approach, such as in echolocation calls of finless porpoises \cite{yang2021transfer}, underwater acoustic recordings \cite{vickers2021robust}, and bird vocalizations \cite{sinha2018deep, zhang2024automatic}.

A common approach for speech enhancement \cite{defossez2020real}, music source separation \cite{jansson2017singing}, speech dereverberation \cite{ernst2018speech}, and denoising animal vocalizations \cite{bulut2020low} uses deep autoencoders, particularly U-Net architectures \cite{ronneberger2015u}. U-Nets are well suited to capturing both global and local features. The U-Net architecture consists of a symmetric U-shaped encoder-decoder. The encoder uses convolution and down-sampling layers to extract features from the input. At each stage, the spatial resolution is reduced, while the feature representation becomes more abstract. In contrast, the decoder restores the original resolution using up-sampling operations. The key feature of the U-Net is the presence of additional skip connections that concatenate the encoder output with the corresponding decoder layers. These skip connections help preserve fine-grained details and improve gradient flow by passing information directly from earlier to later layers, which is crucial for accurately reconstructing the output from the input.

Despite these strengths, U-Net–based denoising approaches rely on supervised training using clean data, which are, however, difficult to obtain in many bioacoustic contexts \cite{stowell2022computational}. To overcome this limitation, we propose a ridge-guided method that leverages ridge information for training set synthesis from noisy recordings and also for weighting the proposed denoiser’s loss function. This approach provides effective denoising without requiring clean training data. In related work, Biodenoising \cite{miron2025biodenoising} produces pseudo-clean training data by applying a pretrained speech-enhancement model or noisereduce \cite{sainburg2024noisereduce} to pre-denoise vocalizations and recombining them with background segments. While both our proposed approach and Biodenoising \cite{miron2025biodenoising} ultimately rely on proxy (pseudo‑clean) targets, the mechanisms for obtaining these targets differ fundamentally. We generate targets analytically from locally sinusoidal components estimated via multicomponent ridge tracking and reconstruct the signal from explicit amplitude, instantaneous‑frequency, and phase parameters. This physics‑informed synthesis avoids dependence on a pretrained denoiser and yields interpretable, controllable targets. Moreover, ridge information is used twice in our framework—both in target synthesis and as a loss weighting that emphasizes ridge-adjacent time–frequency regions—further distinguishing our approach from \cite{miron2025biodenoising}.

Recently, a preprint on training set synthesis for few-shot bioacoustic sound-event detection was posted online \cite{hoffman2025synthetic}. Although related, thar work focuses on generating strongly labeled synthetic scenes, rather than synthesizing clean audio signals suitable for denoising as in the present study.

%%%%%%%%%%%%%%%
%Gaps in Bioacoustic Denoising Approaches
%%%%%%%%%%%%%%%%%
\subsection{Ridge-Tracking in Bioacoustic Processing}\label{sec:relwork_ridge}
Traditional signal-processing methods, such as spectrogram peak tracking and energy-based contour tracking, have been widely used in bioacoustics to extract call contours (ridges) \cite{heller2008automatic, tachibana2020usvseg}. However, these approaches often struggle with noisy calls. To improve robustness, recent studies have applied deep learning approaches for call contour tracking\cite{li2023using, best2025bioacoustic}. These methods typically use ridges extracted by other algorithms \cite{gruden2020automated,warren2022maturation} or manual annotations as training targets to improve the accuracy and robustness of ridge tracking, especially in challenging acoustic environments.

All deep learning approaches, so far, have focused on enhancing the accuracy of ridge-tracking, whereas we take a novel approach: rather than relying on perfectly tracked ridges, we leverage ridge estimates — even if imperfect — to synthesize a training set for a deep learning-based denoiser. This denoising step improves the SNR and makes the data suitable for several downstream analyses such as classification, ridge tracking, and source separation \cite{bermant2021biocppnet}.

\section{Method}\label{sec:method}

%%%%%%%%%%%%%%
%figure 1
%%%%%%%%%%%%%%
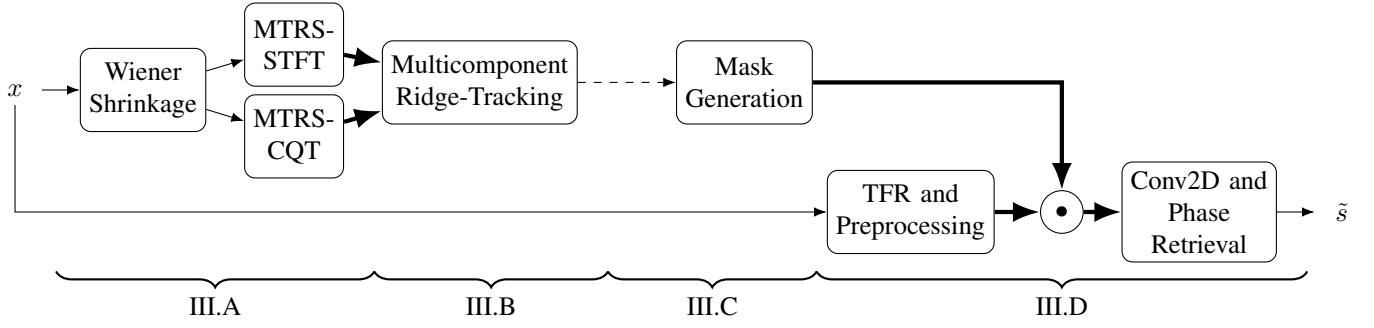
\begin{figure*}[ht]
\centering
\begin{tikzpicture}[
    node distance=0.5cm and 0.5cm,
    box/.style={rectangle, draw, rounded corners, minimum height=1.1cm, minimum width=1.2cm, align=center},
    op/.style={circle, draw, minimum size=0.5cm},
    plain/.style={minimum width=0.7cm, align=center},
    >=Latex
]

% First row
\node (x) [plain] {$x$};
\node (wiener) [box, right=of x] {Wiener \\ Shrinkage};
\node (mtstft) [box, right=of wiener, yshift=0.6cm] {MTRS-\\STFT};
\node (mtcqt)  [box, right=of wiener, yshift=0.4cm, below=of mtstft] {MTRS-\\CQT};

% Second row
\node (ridge) [box, right=of mtstft, yshift=-0.5cm] {Multicomponent\\ Ridge-Tracking};
\node (mask) [box, right=of ridge, xshift=0.8cm] {Mask\\ Generation};

% Third row
\node (tfr) [box, below=0.6cm of mask, xshift=2.2cm] {TFR and \\ Preprocessing};
\node (multiply) [op, right=0.6cm of tfr] {$\bullet$};
\node (phase) [box, right=of multiply] {Conv2D and \\Phase \\Retrieval};
\node (out) [plain, right=of phase, xshift=0cm] {$\tilde{s}$};

% Arrows
\draw[->] (x) -- (wiener);
\draw[->] (wiener) -- (mtstft);
\draw[->] (wiener) -- (mtcqt);

\draw[->, ultra thick] (mtstft) -- (ridge);
\draw[->, ultra thick] (mtcqt) -- (ridge);
\draw[->, ultra thick] (mask) -| (multiply);

\draw[->, dashed] (ridge) -- (mask);
\draw[->] (x) |- (tfr);
\draw[->, ultra thick] (tfr) -- (multiply);
\draw[->, ultra thick] (multiply) -- (phase);
\draw[->] (phase) -- (out);

% Flat braces below the entire diagram
\coordinate (b1start) at ([xshift=0.9cm]x.west);
\coordinate (b1end)   at ([xshift=0.4cm]mtcqt.east);

\coordinate (b2start) at ([xshift=0.4cm]mtstft.east);
\coordinate (b2end)   at ([xshift=0.4cm]ridge.east);

\coordinate (b3start) at ([xshift=0.4cm]ridge.east);
\coordinate (b3end)   at ([xshift=0.05cm]mask.east);

\coordinate (b4start) at ([xshift=0.05cm]mask.east);
\coordinate (b4end)   at ([xshift=0.4cm]phase.east);

\def\bracey{-2.4} % consistent y-position for flat braces

\draw[decorate,decoration={brace,mirror,amplitude=6pt},thick]
    (b1start |- 0,\bracey) -- (b1end |- 0,\bracey) node[midway,below=6pt] {III.A};

\draw[decorate,decoration={brace,mirror,amplitude=6pt},thick]
    (b2start |- 0,\bracey) -- (b2end |- 0,\bracey) node[midway,below=6pt] {III.B};
\draw[decorate,decoration={brace,mirror,amplitude=6pt},thick]
    (b3start |- 0,\bracey) -- (b3end |- 0,\bracey) node[midway,below=6pt] {III.C};

\draw[decorate,decoration={brace,mirror,amplitude=6pt},thick]
    (b4start |- 0,\bracey) -- (b4end |- 0,\bracey) node[midway,below=6pt] { III.D};

\end{tikzpicture}
\caption{Flowchart of the proposed training set synthesis pipeline. The input noisy signal \(x\) is first denoised via Wiener shrinkage to suppress broadband noise. Two time-frequency representations (MTRS-STFT – Multi-Taper Reassigned Short-Time Fourier Transform, MTRS-CQT – Multi-Taper Reassigned Constant-Q Transform) are computed for multicomponent ridge tracking. The ridge-tracking module is not restricted to this specific implementation; alternative ridge-tracking methods and time--frequency parameters may be used depending on the target species. A ridge-guided mask is generated and applied to the time--frequency reassignment (TFR) of \(x\) to isolate signal-dominated regions. The masked TFR is then convolved in 2-D with a Hann window, followed by phase retrieval for training set synthesis \(\tilde{s}\). 
}
\label{fig:synth_pipeline}
\end{figure*}

%%%%%%%%%%%%%%%%%%%%
% figure 2
%%%%%%%%%%%%%%%%%%%%%%
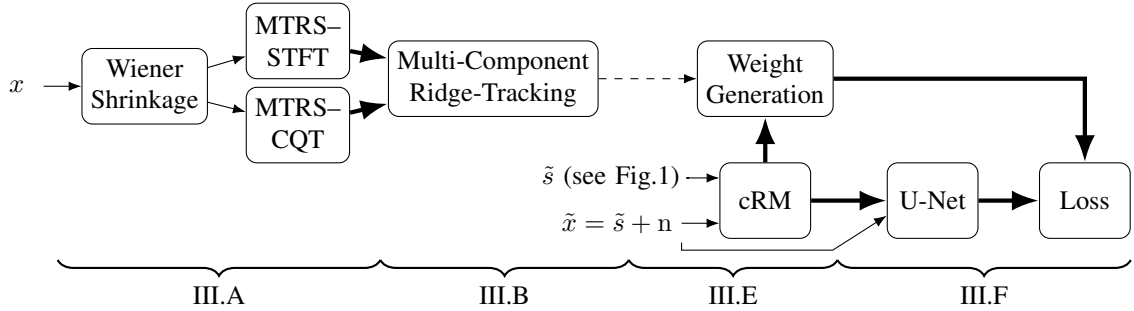
\begin{figure*}[ht]
\centering
\begin{tikzpicture}[
    node distance=0.5cm and 0.5cm,
    box/.style={rectangle, draw, rounded corners, minimum height=1cm, minimum width=1.2cm, align=center},
    op/.style={circle, draw, minimum size=0.5cm},
    plain/.style={minimum width=0.7cm, align=center},
    >=Latex
]

% First row
\node (x) [plain] {$x$};
\node (wiener) [box, right=of x] {Wiener \\ Shrinkage};
\node (mtstft) [box, right=of wiener, yshift=0.6cm] {MTRS--\\STFT};
\node (mtcqt)  [box, right=of wiener, yshift=0.4cm, below=of mtstft] {MTRS--\\CQT};

% Second row
\node (ridge) [box, right=of mtstft, yshift=-0.5cm] {Multi-Component\\ Ridge-Tracking};
\node (mask) [box, right=of ridge, xshift=0.8cm] {Weight\\ Generation};

\node (crm_target) [box, below=0.6cm of mask, xshift=0cm] {cRM};

% Third row
\node (unet) [box, right=0cm of crm_target, xshift=1cm] {U-Net};
%\node (multiply) [op, right=0.8cm of tfr] {$\times$};
\node (xtilde)  [plain, left=0.4cm of crm_target, yshift=0.3cm] {$\tilde{s}$ (see Fig.1)};

\node (snoisy) [plain, left=0.5cm of crm_target, yshift=-0.3cm] {$\tilde{x}=\tilde{s} + \mathrm{n}$};

\node (loss) [box, right=of unet, xshift=0.3cm] {Loss};

% Arrows
\draw[->] (x) -- (wiener);
\draw[->] (wiener) -- (mtstft);
\draw[->] (wiener) -- (mtcqt);

\draw[->, ultra thick] (mtstft) -- (ridge);
\draw[->, ultra thick] (mtcqt) -- (ridge);
\draw[->, ultra thick] (mask) -| (loss);

\draw[->, dashed] (ridge) -- (mask);

\draw[->] (xtilde) -- ++(1cm, 0) -- (crm_target.west |- xtilde);

\draw[->,  ultra thick] (crm_target) -- (mask);
\draw[->] (snoisy) -- (crm_target.west|-snoisy);
\draw[->] 
    (snoisy.south east) 
    -- ++(0,-0.1)         % go down
    -- ++(2,0)            % go right
    -- ++(0,0)          % go up
    -- ([yshift=-0.2cm]unet.west);  % final segment into U-Net
\draw[->, ultra thick] (crm_target.east) -- (unet.west);

\draw[->, ultra thick] (unet) -- (loss);

% Flat braces below the entire diagram
\coordinate (b1start) at ([xshift=0.9cm]x.west);
\coordinate (b1end)   at ([xshift=0.4cm]mtcqt.east);

\coordinate (b2start) at ([xshift=0.4cm]mtstft.east);
\coordinate (b2end)   at ([xshift=0.4cm]ridge.east);

\coordinate (b3start) at ([xshift=0.4cm]ridge.east);
\coordinate (b3end)   at ([xshift=0.05cm]mask.east);

\coordinate (b4start) at ([xshift=0.05cm]mask.east);
\coordinate (b4end)   at ([xshift=0cm]loss.east);

\def\bracey{-2.3} % consistent y-position for flat braces

\draw[decorate,decoration={brace,mirror,amplitude=6pt},thick]
    (b1start |- 0,\bracey) -- (b1end |- 0,\bracey) node[midway,below=6pt] {III.A};

\draw[decorate,decoration={brace,mirror,amplitude=6pt},thick]
    (b2start |- 0,\bracey) -- (b2end |- 0,\bracey) node[midway,below=6pt] {III.B};

\draw[decorate,decoration={brace,mirror,amplitude=6pt},thick]
    (b3start |- 0,\bracey) -- (b3end |- 0,\bracey) node[midway,below=6pt] {III.E};

\draw[decorate,decoration={brace,mirror,amplitude=6pt},thick]
    (b4start |- 0,\bracey) -- (b4end |- 0,\bracey) node[midway,below=6pt] { III.F};

\end{tikzpicture}
\caption{Flowchart of the proposed denoiser training pipeline. The input noisy signal $x$ is first denoised via Wiener shrinkage, and two time--frequency representations (MTRS-STFT and MTRS-CQT) are computed for multicomponent ridge tracking. A weight matrix is generated (Weight Generation) and a complex ratio mask (cRM) target is computed from the synthetic clean signals $\tilde{s}$ and their noisy versions $\tilde{x}=\tilde{s} + \mathrm{n}$ (see Fig.~\ref{fig:synth_pipeline}). The U-Net takes the noisy signal $\tilde{x}$ as input and predicts the cRM. A ridge-guided loss is used, where ridge-based weights emphasize ridge regions in the loss function. The U-Net outputs cRM, which is applied to the noisy input, either \(\tilde{x}\) for synthetic testing data or \(x\) for real testing data, to produce the denoised signal. Brackets below indicate relevant paper sections. 
%Abbreviations: MTRS-STFT – Multi-Taper Reassigned Short-Time Fourier Transform, 
%MTRS-CQT – Multi-Taper Reassigned Constant-Q Transform, 
%U-Net – Convolutional Neural Network with a U-shaped architecture.
}
\label{fig:denoiser_pipeline}
\end{figure*}
%%%%%%%%%%%%%%%%%%
%
%%%%%%%%%%%%%%%%%%
Given a noisy signal \(x\), we assume that it is composed of the original clean bioacoustic signal \(s\) and noise \(n\), such that \(x=s+n\). Because \(s\) is not directly observable, we synthesize a training set to generate an approximate clean signal \(\tilde{s}\) from \(x\). To obtain \(\tilde{s}\), we first track the ridges of the noisy signal \(x\) (Section \ref{sec:ridge_tracking_init}),
using a multicomponent ridge-tracking method. 
Although we use the ridge-tracking approach of~\cite{abbasi2025robust} in this work, the proposed denoising framework is not restricted to a specific ridge-tracking algorithm, and alternative ridge-tracking methods may be incorporated depending on the characteristics of the studied signals.
A mask is then generated based on the obtained ridges (Section \ref{sec:gen_mask}). By applying this mask to the time--frequency reassignment of \( x \) (Section \ref{sec:reass}), the synthetic clean signal \( \tilde{s} \) is constructed following  \cite{rajbamshi2019adhoc} (Section \ref{sec:gen_synth_clean}. The overall pipeline of the training set synthesis is illustrated in Fig.~\ref{fig:synth_pipeline}.

The noisy inputs $\tilde{x}$ are generated by adding noise $n$ to the synthetic data $\tilde{s}$ (Section \ref{sec:gen_noisy_data}), i.e., $\tilde{x} = \tilde{s} + n$. A denoiser network is then trained with the noisy input $\tilde{x}$ and its corresponding clean target $\tilde{s}$. The network predicts complex ratio masks (cRMs) \cite{williamson2015complex} (Section \ref{sec:denoiser}) from noisy spectrograms, which are subsequently applied to obtain the denoised signals. A ridge-guided loss is employed to emphasize vocalization regions by incorporating weights into the loss function (Section \ref{sec:weight}). 
This denoising pipeline is depicted in Fig.~\ref{fig:denoiser_pipeline}. A detailed explanation of each component of our method is provided below.

\subsection{Reassignment}\label{sec:reass}
We use the ridge-tracking method proposed in~\cite{abbasi2025robust} in a case study on USV signals, to estimate the frequency ridge associated with the fundamental frequency and the first harmonic partial. The fundamental frequency is the lowest frequency component of a vocalization, and the harmonic partials occur at integer multiples of this frequency. We note that this definition of ridge-tracking is valid for species in which the fundamental frequency corresponds to the component with maximum amplitude; for other species, such as elephants \cite{stoeger2014age}, the method may need to be adjusted to identify the fundamental frequency, in order to subsequently estimate correct harmonics.

Following \cite{abbasi2025robust}, the first step in ridge tracking is to reduce stationary noise in the signal \(x\). This is achieved by applying empirical Wiener shrinkage~\cite{ghael1997improved} on the coefficients of a short-time Fourier transform (STFT) \cite{allen1977short}. The noise power spectral density is estimated from the spectrogram of a 2-second noise-only segment containing no bioacoustic signal. We refer to the result of this step as the Wiener-denoised signal. Then, we compute two time-frequency representations of the Wiener-denoised signal for different processing steps: an STFT to track the fundamental ridge and a constant-Q transform (CQT)~\cite{brown1991calculation} to track the harmonic-partial ridge.

The time-frequency resolution of both STFT and CQT is constrained by the uncertainty principle~\cite{grochenig2003uncertainty}, resulting in unavoidable smearing in their spectrograms. Spectrogram reassignment \(R_x\) sharpens these representations by relocating energy based on local instantaneous frequency and group delay~\cite{aufl95, holighaus2016reassignment}. To further attenuate noise and improve robustness, we use multitaper spectrogram reassignment (MTRS-STFT)~\cite{xiao2007multitaper}, extended to the CQT (MTRS-CQT) based on~\cite{holighaus2016reassignment}. This is done by averaging reassigned spectrograms obtained with orthogonal windows (e.g., Hermite functions), emphasizing deterministic signal components while suppressing background noise. We further refine the MTRS-CQT using a modified version of the pre-denoising step in~\cite[Section III-B]{xiao2007multitaper}, where we replace the binary mask with a smooth, continuous weighting function that softly attenuates weak components and avoids hard thresholding.

\subsection{Multicomponent Tracking}\label{sec:ridge_tracking_init}

Following ~\cite{abbasi2025robust}, the fundamental ridge is identified in the MTRS-STFT via path optimization using MATLAB’s \texttt{tfridge}, which tracks the maximum-energy path while minimizing abrupt jumps between consecutive time steps. 

To detect harmonic partials, we leverage the fact that harmonic partials occur at constant distances from the fundamental frequency in the CQT. For each MTRS-CQT time frame, we compute the autocorrelation of the frequency vector to detect harmonic partials. In this study, the analysis is restricted to the first harmonic partial, since mouse USVs typically occupy high-frequency ranges (e.g., fundamental frequencies around 50--60\,kHz), making higher harmonics less consistently observable within the analyzed frequency range. To enhance robustness against noise and spurious peaks, we apply a weighted average of the autocorrelations over three adjacent time frames. The CQT’s high frequency resolution ensures that each bin captures a quasi-sinusoidal partial, resulting in a clearer dominant peak for reliable harmonic detection. 

Importantly, the MTRS-STFT and MTRS-CQT are used only for multicomponent ridge tracking. For the other parts of the proposed method, we rely on the STFT coefficients.

\subsection{Mask Generation for Training Set synthesis}\label{sec:gen_mask}
In this study, we adopt the approach proposed in \cite{rajbamshi2019adhoc} to generate clean synthetic signals \(\tilde{s}\), which is based on inverting the reassigned spectrogram \(R_x\). 
To emphasize the spectral content around the fundamental and the harmonic-partial ridges while suppressing background noise, we multiply \(R_x\) pointwise with truncated Gaussian masks. We define the masking matrix \(G \in \mathbb{R}^{N \times K}\), where \(N\) is the number of time frames and \(K\) is the number of frequency bins.

For each time frame \(n\), we construct a Gaussian-shaped mask vector centered at the ridge frequency \(f_r[n]\), with nonzero values only within a window of length \(L\) frequency bins around the ridge. Specifically, the mask at time \(n\) is defined as
\begin{equation}
\begin{aligned}
G[n, k] &= \exp\left(-\frac{(k - k_r[n])^2}{2\sigma^2}\right), \\
k &\in \left\{ k_r[n] - \frac{L}{2}, \dots, k_r[n], \dots, k_r[n] + \frac{L}{2} \right\}.
\end{aligned}
\label{eq:mask}
\end{equation}
where \(k\) is the frequency bin index, \(k_r[n]\) is the bin corresponding to the ridge frequency \(f_r[n]\) at time frame \(n\), and \(\sigma = L/4\) controls the spread of the Gaussian mask. Outside this interval, \(G[n,k]\) is set to zero. The mask is normalized to have a maximum value of 1.

Using the reassigned spectrogram \(R_x\) and the Gaussian mask \(G\), we construct a masked reassigned spectrogram \(R_x^\text{masked}\). This is obtained as \(R_x^\text{masked}[n, k] = G[n, k] \cdot
R_x[n, k]\).
%%%%%%%%%%%%%%%%%%%
%Audio Synthesis
%%%%%%%%%%%%%%%%%%%
\subsection{Audio Synthesis}\label{sec:gen_synth_clean}

The approach of \cite{rajbamshi2019adhoc} consists of two main steps. First, the reassigned spectrogram \(R_x\) coefficients are convolved with a kernel \(W\) to approximate the original STFT magnitude spectrum \(C_x\). The kernel \(W\) is constructed from the Hann window \(w_h[n]\) and its discrete Fourier transform \(\hat{w}_h[k]\). The convolution is given by:

\begin{equation}
\begin{aligned}
C_x[n,k] &= (R_x \ast W)[n,k] \\
&= \sum_{n'} \sum_{k'} R_x[n',k'] \, W[n - n', k -k'], \\
\text{where} \quad W[n,k] &= w_h[n] \otimes \hat{w}_h[k].
\end{aligned}
\label{eq:inv_reass_combined}
\end{equation}

Second, the phase is estimated via the fast Griffin-Lim algorithm \cite{perraudin2013fast}, and the synthetic clean signal \(\tilde{s}\) is recovered via inverse STFT.
In our work, we modify this procedure by using the masked reassigned spectrogram \(R_x^{\text{masked}}\) in place of \(R_x\) in Eq.~\eqref{eq:inv_reass_combined}, allowing us to emphasize spectral components around the ridges while suppressing noise and other components.

\subsection{Weights for the Ridge-Guided Loss Function}
\label{sec:weight}
For training the proposed denoiser, we use the synthetic clean signal \(\tilde{s}\) and the corresponding noisy signal \(\tilde{x}\), which is obtained by adding noise to \(\tilde{s}\) (see Section~\ref{sec:denoiser}). The denoiser takes the spectrogram of the noisy signal as input and predicts a cRM \cite{williamson2015complex} as the output. The cRM is a time--frequency mask that operates directly on the complex spectrogram. Unlike magnitude-only masks, the cRM accounts for both magnitude and phase information, enabling more accurate reconstruction of the denoised signal.
Let \( S_{\tilde{x}} \) and \(S_{\tilde{s}} \) denote the complex spectrograms of the noisy and clean signals, respectively. 
The ideal mask \(M_x\), obtained from the training dataset, is defined as
\begin{equation}
    \mathrm{M_x} = \frac{S_{\tilde{s}}}{S_{\tilde{x}} + \varepsilon},
    \label{eq:crm}
\end{equation}
where \(\varepsilon = 10^{-8} \) is a small constant added for numerical stability.

To improve training robustness and bound the dynamic range, a smooth compressive nonlinearity is applied separately to the real and imaginary components of \(M_x\). The resulting bounded mask, denoted \(O_x\), is computed as
\begin{equation}
     O_x= B \, \frac{1 - e^{-C.M_x}}{1 + e^{-C.M_x}},
    \label{eq:compressed_crm}
\end{equation}
where \( B \) is a scale factor and \( C \) is a slope parameter. 
This transformation compresses \(M_x\) values into the range \(( -B, B )\). In our experiments, we set \( B = 15 \) and \( C = 0.1 \).

The complex-valued mask \(O_x\) can be written as \(O_x = O_{x_\mathrm{r}} + j\, O_{x_\mathrm{i}}\), where \(O_{x_\mathrm{r}}\) and \(O_{x_\mathrm{i}}\) denote the real and imaginary components, respectively, and are used as the training targets for the proposed denoiser.

While the \(O_x\) provides a representation of the signal’s structure, all time-frequency regions are treated equally during training. 
However, not all regions contribute equally to the accurate signal reconstruction, as ridges carry more information than background noise. To address this, we design a weighting matrix that emphasizes ridge regions during loss computation (Section \ref{sec:denoiser}), computed separately for the real and imaginary targets \(O_{x_\mathrm{r}}\) and \(O_{x_\mathrm{i}}\).

For each time frame \(n\), weights are assigned within a band of \(\pm b\) bins centered at the fundamental and harmonic partial ridge positions. The weighting is computed separately for the real \(O_{x_\mathrm{r}}\) and imaginary \(O_{x_\mathrm{i}}\) components of the \(O_x\), yielding weight matrices \(w_\mathrm{r}[n,k]\) and \(w_\mathrm{i}[n,k]\). We define $w_\mathrm{r}[n,k]$ as
% \begin{equation}
%     w_r[n, k] = \gamma \left[ \alpha + (1 - \alpha) \left( 1 - \frac{O_{x_\text{r}}[n, k]}{O_{x_\text{r}, \text{max}} } \right) \right],
%     \label{eq:loss_weights}
% \end{equation}

\begin{equation}
\label{eq:loss_weights}
\begin{aligned}[t]
w_r[n,k] &=
\begin{cases}
\gamma \Big[ \alpha + (1 - \alpha)
\Big( 1 - \frac{O_{x_r}[n,k']}{O_{x_r,\max}} \Big) \Big],
 \\[-0.1ex] \hspace{8em}\text{if } |k - k'| \le b, \\
1, & \text{otherwise}.
\end{cases}
\end{aligned}
\end{equation}

Here, $O_{x_\mathrm{r}, \max}$ is the maximum of the matrix $O_{x_\mathrm{r}}$, 
$\alpha = 0.3$ is a floor factor that prevents the ridge weight from becoming too small, 
and $\gamma = 10$ is a scaling constant controlling the overall ridge weight. 
The weights $w_\mathrm{i}$ are computed analogously for the imaginary component, 
using $O_{x_\mathrm{i}}$ in place of $O_{x_\mathrm{r}}$, allowing the weighting to adapt independently to the local structure of both components of the complex representation.
The resulting weights $w_\mathrm{r}$ and $w_\mathrm{i}$ are clipped to the range \([6.0, 10.0]\) to prevent extreme weighting that could destabilize training. Lower-amplitude ridge regions receive larger weights because faint vocalization components are more susceptible to suppression during denoising than high-energy ridge regions, which are typically easier to preserve.
These weighting matrices are applied to their corresponding loss terms during denoiser training (Section~\ref{sec:denoiser}).

\subsection{Denoiser}
\label{sec:denoiser}
Given the need to preserve the local features and reconstruct the fine-grained spectral structures, we adopt a U-Net \cite{ronneberger2015u} architecture for denoising. In this study, the network operates on the magnitude spectrogram of the noisy signal \(\tilde{x}\), which captures changes in both amplitude and frequency over time, and predicts the mask \(\hat{O}_x\). So, training is performed in a supervised manner, with \(O_x\) (see Eq.~\eqref{eq:compressed_crm}) used as the U-Net target. To guide the U-Net toward preserving ridges while suppressing residual noise, we employ a ridge-guided loss. The total loss \(\mathcal{L}_\text{total}\) (see Eq.~\eqref{eq:loss_total}) accounts for both the real (\(\hat{O}_{x_\mathrm{r}}\)) and the imaginary (\(\hat{O}_{x_\mathrm{i}}\)) components of the predicted mask \(\hat{O}_x\) and includes a penalty term for non-ridge regions to reduce undesired background energy.

Let \(\hat{O}_{x_\mathrm{r}}\) and \(\hat{O}_{x_\text{i}}\) denote the real and imaginary components of the mask predicted by the U-Net.  
Using weights \(w_\mathrm{r}[n,k]\) and \(w_\mathrm{i}[n,k]\) (see Eq.~\eqref{eq:loss_weights}), the weighted loss \(\mathcal{L}_{O_x}\) is defined as
\begin{equation}
\begin{aligned}
\mathcal{L}_{O_x} =   \\
\frac{1}{N K} \sum_{n=1}^{N} \sum_{k=1}^{K} 
& \Bigl(w_\mathrm{r}[n,k] \, \big(O_{x_\mathrm{r}}[n,k] - \hat{O}_{x_\mathrm{r}}[n,k]\big)^2 \\
+ & w_\mathrm{i}[n,k] \, \big(O_{x_\mathrm{i}}[n,k] - \hat{O}_{x_\mathrm{i}}[n,k]\big)^2\Bigr),
\label{eq:loss_complex}
\end{aligned}
\end{equation}

In addition, a penalty term \(\mathcal{L}_\mathrm{noise}\) is included to suppress energy in the predicted mask for non-ridge (noise) regions. 
This term is weighted by the magnitude of the STFT of the noisy input \(\tilde{x}\), and is computed as

\begin{equation}
\mathcal{L}_\text{noise} = 
\frac{1}{N K} \sum_{n,k} 
\mathbf{1}_{\{w_\mathrm{r}[n,k] = 1\}} \, \big|\tilde{X}[n,k]\big| \, \big|\hat{O}_x[n,k]\big|^2,
\label{eq:loss_noise}
\end{equation}

where $|\hat{O}_x[n,k]|$ is the magnitude of the predicted mask \(\hat{O}_x\), $|\tilde{X}[n,k]|$ is the magnitude of the STFT of the noisy input  $\tilde{x}$, and $\mathbf{1}_{\{w_\mathrm{r}[n,k] = 1\}}$ is the indicator function selecting the time–frequency bins corresponding to noise regions.

The ridge-guided loss used for training the denoiser combines these two terms from  Eq.~\eqref{eq:loss_complex} and  Eq.~\eqref{eq:loss_noise}:
\begin{equation}
\mathcal{L}_\text{total} = \mathcal{L}_{O_x} + \alpha_\text{noise} \, \mathcal{L}_\text{noise},
\label{eq:loss_total}
\end{equation}
where \(\alpha_\text{noise}\) controls the relative importance of the noise penalty. In our experiments, we set \(\alpha_\text{noise} = 0.8\) based on pilot tasks. 

After training, the denoiser predicts the bounded mask $\hat{O}_x$ 
in the compressed domain. However, the uncompressed mask is required to 
obtain the complex STFT of the denoised signal (see Eq.~\eqref{eq:crm}). 
Therefore, an estimate of the uncompressed mask $\hat{M}_x$ is recovered 
from the predicted compressed mask $\hat{O}_x$ by applying the inverse of 
the compressive nonlinearity:

\begin{equation}
    \hat{M}_x = -\frac{1}{C} \log \left( \frac{B - \hat{O}_x}{B + \hat{O}_x} \right).
    \label{eq:uncompressed_crm}
\end{equation}

The recovered uncompressed mask $\hat{M}_x$ is then used to reconstruct 
the complex STFT of the denoised signal, which serves as input for subsequent analyses.

\section{Experimental Setup}\label{sec:experiment}
This section describes the mouse USV dataset and pre-processing, implementation details, evaluation methods used to assess the proposed denoiser’s impact on the SI-SDR and ridge-tracking performance, as well as the classification accuracy achieved by the integrated denoising-and-retraining pipeline. 

\subsection{Dataset and Pre-processing}\label{ssec:data}
The USVs of adult wild-derived house mice (\textit{Mus musculus musculus}) used in the present study were previously recorded and described in \cite{klaus2025courtship}. The vocalizations were recorded from adult mice using the RECORDER USGH software (Avisoft-RECORDER Version 4.2) at a sampling rate of 300 kHz and a 16-bit format. Fundamental frequencies range from 40 to 120 kHz, with durations between approximately 5-200 ms. 

Biologists experienced in annotating mouse vocalizations classified the USVs into 11 distinct categories (or syllable types) based on several acoustic features, as described in \cite{klaus2025courtship}, i.e., upward (“UP"), downward (“D"), “C2" for two-component USVs, “C3" for three-component USVs, complex (“C"), U-shaped and inverted U-shaped USVs labeled as “U" and “UI", and flat (“F"), short (“S"), ultrashort (“US"), and all USVs with harmonic partials are denoted as “H". 
Because there are very few samples in the “US" and “S" classes, they are excluded from further analysis. Although the original study annotated some USVs with harmonic components, here we use labels based solely on the primary syllable type, ignoring harmonic features. USVs were labeled using spectrograms (visual inspection) and the Automatic Mouse Ultrasound Detector (A-MUD) for automatic detection \cite{Zala2017, zala2020primed}. 
To preserve the full extent of each vocalization, USV segments are extended by $\pm3$ ms beyond the time intervals detected by A-MUD. This extension is chosen empirically after testing $\pm0.5$ ms, $\pm1.5$ ms, $\pm3$ ms, and $\pm5$ ms; visual inspection confirm that the extended segments did not include wideband bursts at their boundaries.

Most existing ridge-tracking methods, including DeepSqueak’s ridge-tracker \cite{coffey2019deepsqueak}, perform well under high SNR conditions, but are sensitive to noise and typically track only the fundamental frequency of each vocalization. Consequently, they neglect harmonic partials, which are mainly emitted during the onset of copulatory behavior in mice~\cite{hanson2012female,klaus2025courtship}. This limitation highlights the need for ridge-tracking techniques that can capture both the fundamental and the harmonic components.  
Therefore, the ridges of the USVs analyzed in the present study are tracked using the multicomponent tracking method described in Section~\ref{sec:ridge_tracking_init}.

We have developed a Python-based graphical user interface (GUI) to assign one of three labels (correct, incorrect, or invalid) to each USV ridge, with each assignment verified by visual inspection. Even for clean USVs, where manual ridge annotation is relatively quick, this GUI reduces the annotation time by at least a factor of ten. In the dataset tested, 83.7\% of USVs have correctly tracked ridges, 13.7\% show incorrectly tracked ridges, and 2.6\% are marked as invalid. The USVs with correctly tracked ridges are used for training, validation, and testing of the proposed denoiser and the classifier, employing the test set for the in-sample evaluation. For training the proposed denoiser, we use the synthesized version of our training data, as described in Section~\ref{sec:implementation}. The subset of incorrectly-tracked USVs is used as out-of-sample testing data for the proposed denoiser and classifier. Invalid samples are excluded from all stages of training and evaluation, as they either contain no detectable USVs or are suspected to include overlapping vocalizations from two mice. A total of 12{,}937 USVs from the “correct" and “incorrect" subsets are analyzed, with sample counts for each class shown in Table~\ref{tab:data_dist}.

\begin{table*}[t]
\centering
\small
\caption{Number of USVs in each class (C, C2, C3, D, F, U, UI, UP), across the training, validation, and testing subsets of the correct-ridge dataset, and the testing subset of the incorrect-ridge dataset. “Subset” indicates the data split used for training or evaluation of the proposed denoiser and classifier.}
\label{tab:data_dist}
\begin{tabular}{@{}lrrrrrrrr@{}}
\hline
       & \multicolumn{8}{c}{\textbf{USV Class}} \\
\cline{2-9}
Subset & C & C2 & C3 & D & F  & U & UI & UP \\
\hline
\multicolumn{9}{c}{\textbf{Correct-Ridge Data}} \\
\hline
Training   & 829 & 262 & 689 & 1205 & 1349 &  534 & 1676 & 2397 \\
Validation & 100 & 33  & 87  & 168  & 163  &  65  & 231  & 272  \\
Testing    & 113 & 27  & 80  & 151  & 156  &  76  & 214  & 299  \\
\hline
\multicolumn{9}{c}{\textbf{Incorrect-Ridge Data}} \\
\hline
Testing  & 194 & 78  & 198 & 265  & 308  &  104 & 239  & 375  \\
\hline
\end{tabular}
\end{table*}

\subsection{Implementation Details}\label{sec:implementation}
\subsubsection{Training Set Synthesis}\label{sec:imp_syn_clean}
To track the ridges introduced in Section~\ref{sec:ridge_tracking_init}, we first denoise \(x\) and then compute the STFT and CQT. 
The STFT uses a Hann window of 750 samples (2.5 ms) and a hop size of 150 samples (0.5 ms), and an FFT size (NFFT) of 750. Similar parameter ranges are commonly used in mouse USV analysis and ridge-tracking studies~\cite{coffey2019deepsqueak,abbasi2022capturing,Zala2017}. 
For the CQT, we set $a = 150$ (0.5 ms hop size), $\xi_{\textrm{min}} = 28$\,kHz (minimum frequency), $B = 512$ (number of frequency bins per octave), 
and $K = 1240$ (total number of frequency bins), covering the frequency range from $\xi_{\textrm{min}}$ up to the Nyquist frequency of 150\,kHz. The relatively high CQT resolution (\(B = 512\)) is selected to improve localization of closely spaced high-frequency ridge structures and harmonic partials in mouse USVs. The number of tapers for both TFRs is set to six to balance noise suppression and spectral concentration without overly smoothing of rapidly modulated vocalizations. 

For fundamental ridge tracking, the penalty parameter is set to $2 \times 10^{-8}$. 
After ridge detection, points with a frequency bin distance exceeding 10 are set to NaN, 
and missing values are linearly interpolated to smooth the ridges. For harmonic partial tracking, after applying autocorrelation on the MTRS-CQT, candidate peaks are restricted to a $\pm 3$\,kHz interval around twice the fundamental frequency, to account for estimation errors, and 60 edge bins at both band limits are zeroed to avoid boundary artifacts. To ensure temporal consistency, only sequences with at least five consecutive valid peaks are retained. Finally, the global maximum within the defined search region is selected as the harmonic partial candidate. 

For computing $R_x$, all audio samples are processed using a fixed 200~ms window with each USV centered in the window. 
The corresponding mask $G$ (Section~\ref{sec:gen_mask}) is computed over the same interval, 
it is nonzero within a window of length \(L=5\) frequency bins around the ridge at each time frame, and zero elsewhere. The mask width is chosen to emphasize ridge neighborhoods while limiting leakage into surrounding noise regions.
The clean training data $\tilde{s}$ are then synthesized following the approach described in Section~\ref{sec:gen_synth_clean}.

\subsubsection{Generating Noisy Data}\label{sec:gen_noisy_data}

For training and validation of the proposed denoiser, noisy samples $\tilde{x}$ are generated from synthesized clean signals $\tilde{s}$ (each 200 ms in duration), such that $\tilde{x}=\tilde{s}+\mathrm{n}$, where noise is generated based on samples drawn from a noise dataset \(N\) and subsequently processed before addition. For the training and validation sets, this noise dataset \(N\) consists of 3{,}240 noise segments (200 ms each) extracted automatically, without manual supervision, from five recordings of three male mice. The extraction is performed from non-vocalization regions identified using the USV annotations together with an energy-based criterion to exclude silent segments. These five recordings contained only 235 USVs out of the complete dataset reported in Table~\ref{tab:data_dist}, providing substantial background-only regions for noise estimation. Additionally, 30 broadband-noise segments (200 ms each)—a subset of segments containing broadband spectral content— are included to simulate the challenging scenarios encountered in actual field recordings.

For each clean synthetic sample \(\tilde{s}\), a target SNR is drawn from one of five bins: (-10, -5), (-5, 0), (0, 5), (5, 10), and (10, 15) dB, with corresponding selection probabilities of 0.30, 0.25, 0.20, 0.15, and 0.10, biasing the training and validation data toward lower SNRs. To construct the noise, between one and three samples are randomly selected from \(N\). With a probability of 0.5, one of these selected noise samples could be replaced by broadband noise. This broadband noise is generated by randomly selecting between one and three broadband noise segments and averaging them. All selected noise segments are then averaged to produce the resulting noise signal.

Finally, the resulting noise is scaled based on root-mean-square (RMS) amplitude to achieve the target SNR relative to the synthetic clean sample. The scaled noise is then added to the synthesized clean signal $\tilde{s}$ to produce the noisy sample $\tilde{x}$.

\subsubsection{U-Net Architecture}
The U-Net we use here consists of four encoder blocks and four decoder blocks, each containing two convolutional layers, for a total of eight convolutional layers in the encoder and eight in the decoder. The number of channels in the encoder starts at 8 and doubles after each downsampling step (8, 16, 32, 64), while in the decoder it is halved symmetrically. Each convolution uses $3\times3$ kernels with stride 1, followed by LeakyReLU activations ($\alpha=0.1$) and Batch Normalization. Downsampling in the encoder is performed using $2\times2$ max-pooling layers with stride 2, followed by dropout with rate 0.3 to reduce overfitting, while upsampling in the decoder uses $2\times2$ transposed convolutions. Skip connections concatenate encoder feature maps with corresponding decoder feature maps to preserve spatial detail. The network includes a bottleneck stage at the lowest resolution with two convolutional layers (128 channels). The output layer predicts a two-channel complex mask (real and imaginary) for the input spectrogram.

\subsubsection{U-Net Training}
For training the proposed denoiser, synthetic clean and noisy signals, $\tilde{s}$ and $\tilde{x}$, are normalized by the RMS of the noisy signal $\tilde{x}$.  
The normalized noisy signal is then transformed into the logarithmically compressed magnitude of its STFT, $\log(1 + |S_{\text{noisy}}|)$.  
The ideal mask is obtained using \eqref{eq:crm} and further compressed according to \eqref{eq:compressed_crm}.

The U-Net model is trained with the Adam optimizer, configured with an initial learning rate of \(10^{-4}\), \(\beta_1 = 0.5\), and \(\beta_2 = 0.9\). The network weights are initialized randomly and the training is conducted using a batch size of 32. To prevent overfitting, early stopping is employed by monitoring the validation loss, with a patience of 5 epochs. The training procedure is configured to restore the best-performing model weights based on the lowest observed validation loss.

After training, the predicted cRM is uncompressed according to Eq.~\eqref{eq:uncompressed_crm} and applied to the complex STFT of the noisy input signal.  
The denoised waveform is then reconstructed via inverse STFT and rescaled by the RMS of the input noisy signal.

\subsubsection{Software and Libraries}
For STFT calculations, we use \texttt{librosa} (v0.8.0) \cite{mcfee2015librosa}. 
For CQT calculations, we use implementations provided by the LTFAT package \cite{pruuvsa2014large,sondergaard2012linear} in MATLAB, based on \cite{dogrhove13, holighaus2016reassignment}.
The U-Net denoiser and classifier are implemented with TensorFlow/Keras (v2.5.0) and trained on GPU-enabled systems equipped with NVIDIA A100-SXM4 (40~GB) GPUs on the CLIP Batch Environment (CBE) cluster and NVIDIA GeForce GTX 1080~Ti GPUs on the Acoustic Research Institute’s cluster.
Data pre-processing and analyses are performed in Python~3.8 using NumPy~1.19.5, and all figures are produced with Matplotlib~3.3.4.

The implementations for the training set synthesis, denoiser and classifier training, the Python-based GUI for ridge evaluation, and the manually annotated dataset containing ridge annotations for 193 USVs are publicly available at \footnote{\url{https://github.com/ReyhanehAbbasi/bioacoustic-denoising}}.

\subsection{Evaluation}
We evaluate the proposed approach using three strategies.  
First, we synthesize a test set to quantify improvements in the scale‑invariant signal‑to‑distortion ratio (SI-SDR) \cite{le2019sdr}.  
Second, we evaluate the effect of denoising on USV ridge-tracking accuracy.  
Third, we assess classification F1-score for in-sample and two out-of-sample test sets, using a classifier retrained on denoised data.
The details of each evaluation strategy are provided below.

\paragraph*{Denoising Methods Under Comparison}
We compare the proposed denoiser with three baseline methods: (1) an analytical denoiser, (2) the noisereduce package \cite{sainburg2024noisereduce} evaluated in both stationary and non-stationary modes, and (3) the pretrained Biodenoising framework \cite{miron2025biodenoising}.

\textbf{Analytical denoiser}: As an analytical baseline, we implement the same ridge-guided signal model used for training-set synthesis, as described in Sections~\ref{sec:reass}--\ref{sec:gen_synth_clean}. Parameters are tuned on a small subset of the lowest-SNR validation data (50 samples at -10 dB) to reduce erroneous ridge-tracking jumps under highly noisy conditions. The analytical denoiser is evaluated for SI-SDR, ridge-tracking accuracy, and USV classification.

\textbf{Noisereduce:}  
As a conventional cross-domain spectral-gating baseline, the Python package noisereduce (v4.0.0) is evaluated in both stationary and non-stationary modes without providing parallel noise-only references, consistent with our test setup. For each mode, a single parameter configuration is selected on a small subset of the lowest-SNR validation data (50 samples at -10 dB) and then held fixed across all SNR levels.

For the stationary mode, we set $n_{\mathrm{thresh}}=2.4$, $f_{\mathrm{mask,smooth}}=800 \,\mathrm{Hz}$, and $t_{\mathrm{mask,smooth}}=5 \,\mathrm{ms}$. %$\mathrm{n\_std\_thresh\_stationary} = 2.4$, $\mathrm{freq\_mask\_smooth\_hz} = 800$, and $\mathrm{time\_mask\_smooth\_ms} = 5$. 
For the non-stationary mode we set $s_{\mathrm{slope}}=12$, $t_{\mathrm{nmult}}=1.5$, $p_{\mathrm{decrease}}=0.9$, $f_{\mathrm{mask, smooth}}=800 \,\mathrm{Hz}$, $t_{\mathrm{mask, smooth}}=3 \,\mathrm{ms}$, and $t_{\mathrm{constant}}=0.02 \,\mathrm{s}$. %$\mathrm{sigmoid\_slope\_nonstationary} = 12$, $\mathrm{thresh\_n\_mult\_nonstationary} = 1.5$, $\mathrm{prop\_decrease} = 0.9$, $\mathrm{freq\_mask\_smooth\_hz} = 800$, $\mathrm{time\_mask\_smooth\_ms} = 3$, and $\mathrm{time\_constant\_s} = 0.02$. 
These configurations are selected conservatively to provide competitive denoising while minimizing excessive suppression of weak USV structure. Both noisereduce variants are evaluated only for SI-SDR.

\textbf{Biodenoising:}  
As the third baseline, we evaluate Biodenoising \cite{miron2025biodenoising} using its publicly available pretrained model, without retraining or domain-specific fine-tuning. Biodenoising operates at 16~kHz, whereas our ultrasonic recordings are sampled at 300~kHz. We rewrote recordings at 16~kHz without waveform resampling, thereby shifting the ultrasonic content into the model’s operating range while preserving waveform sample structure. After denoising, the outputs are restored to the original ultrasonic timescale by writing the processed waveform back at the original 300~kHz sample rate. This baseline therefore provides a cross-domain pretrained benchmark rather than a USV-specialized denoiser. Biodenoising is evaluated for SI-SDR comparisons only.

\paragraph*{Scale-Invariant Signal-to-Distortion Ratio (SI–SDR)}
Because original recordings lack clean reference signals, we synthesize a test set to compute improvements in the SI-SDR. To generate this test set, we use 
ridges from the testing subset of the correct-ridge data described in Table~\ref{tab:data_dist}. We then follow the same procedure as described in Section~\ref{sec:gen_synth_clean} for the synthesis of the testing set $\tilde{s}$.  
Then, noisy testing data \(\tilde{x}\) are produced following the approach described in Section~\ref{sec:gen_noisy_data}.  
However, the noisy testing data differ from the noisy training data in several aspects: (i) fixed SNRs of (-10, -5, 0, 5, 10, and 15) dB are used, instead of randomly sampled SNRs, (ii) broadband noise segments are excluded, and (iii) the noise samples are obtained from two recordings different from those used for noisy training-data generation, containing 159 noise segments. 
As this test set is generated using a related synthesis procedure, SI-SDR results should be interpreted as a controlled, relative comparison under synthetic conditions.

\paragraph*{Accuracy of Tracked Ridges}
We assess the impact of denoising on the accuracy of ridge-tracking both with and without weighting the cRM components in the denoiser's loss function (Eq.~\eqref{eq:loss_complex}). When data are denoised using the U-Net denoiser, the ridge-tracking approach in Section \ref{sec:ridge_tracking_init} is applied directly on the denoised data. The same parameters asin \ref{sec:imp_syn_clean} are used for the fundamental ridge-tracking penalization, ridge-modification, and harmonic partial ridge-tracking. For comparison, we also report the ridge-tracking results obtained using the baseline methods \cite{abbasi2025robust, coffey2019deepsqueak}. The first method corresponds to the pipeline described in Sections \ref{sec:reass} and \ref{sec:ridge_tracking_init}, while the second method is the approach implemented in DeepSqueak, a commonly used tool for analyzing USVs.

All methods are evaluated against manually annotated fundamental and harmonic partial ridges, using the same dataset described in \cite{abbasi2025robust}. In this previous study, the ridges were initially estimated with \cite{abbasi2025robust} and subsequently they are manually refined post hoc using a custom Python GUI. These samples cover a range of low- and high-SNR cases.

For evaluation, the estimated fundamental and harmonic partial ridges for each USV are compared pointwise with manually annotated ridges. Performance is measured using precision and mean frequency deviations. An estimated ridge is considered a true positive if its difference from the manually annotated ridge is within half a semitone (one-twelfth of an octave) \cite{salamon2014melody,best2025bioacoustic}; otherwise, it is treated as a false positive.
The same criterion is applied to harmonic partial ridges. False positives can occur whenever no first harmonic partial is annotated, and false negatives whenever a harmonic partial is present but missed. Therefore, recall is also reported for harmonic partial ridges. However, recall is not measured for the fundamental ridge, as the ridge-tracking method we use here tracks ridges for all time bins in the input data, resulting in no false negatives. Because  DeepSqueak does not estimate harmonic partials, harmonic partial metrics are not reported for that method.
%%%%%%%%%%%%%%%%%%
%Denoising Effect on USV Classifier Performance in Moderately and Heavily Noisy Conditions
%%%%%%%%%%%%%%%%%%
\paragraph*{F1-score of the USV Classifier}\label{sec:eval_classification}
We assess the effect of the proposed integrated pipeline on the performance of the USV classifier. %Here, we assess the effect of the proposed denoiser on the performance of the USV classifier.
To classify USVs, we train the BootSnap model described in \cite{abbasi2022capturing}. BootSnap is a five-layer CNN classifier trained on bootstrapped training data. We train three separate BootSnap models on the training subset of the correct-ridge dataset (see Table~\ref{tab:data_dist}). One classifier is trained on this subset using the original audio,   
%The second model is trained on denoised USVs using the proposed denoising approach.
the second classifier is trained on the same subset after applying the analytical denoiser, and the third classifier is trained after applying the proposed denoiser. In all cases, BootSnap’s built‑in default denoising is applied during training to maintain pipeline compatibility and each audio signal is peak-normalized before being fed into the BootSnap pipeline.

To assess the performance of the BootSnap models, we evaluate the three classifiers on three test datasets:
1) The test subset the correct-ridge data (Table~\ref{tab:data_dist}), which contains high-SNR USVs similar to the correct-ridge training data used to train the BootSnap models; 2) the incorrect-ridge subset (Table~\ref{tab:data_dist}), which consists primarily of low-SNR USVs with a higher proportion of faint and noise-overlapped USVs, making them more difficult to denoise and classify; and 3) an out‑of‑sample set of 2,223 USVs from a domesticated laboratory mouse strain (B6D2F1, recorded for \cite{chabout2015male}), using fresh urine as stimulus, as previously described in Table 1 of \cite{abbasi2022capturing}. %This set is used to assess the effect of the denoiser on the generalizability of the classifier
This third set evaluates the robustness of the proposed integrated pipeline to a domesticated strain, which is genetically and behaviorally distinct from wild mice.
 
The third dataset was originally used in \cite{abbasi2022capturing} to assess the generalizability of the BootSnap. It is important to note that the original labeling of this dataset consists of 5 classes instead of 8. Therefore, to be able to compare both datasets to make predictions, we pool the syllab “Rise", yielding 5 output classes (i.e., Rise, UI, C, C2, and C3). 
We emphasize that these experiments evaluate the integrated pipeline (denoising and classifier retraining). Attributing gains to a denoiser alone would require a cross‑condition test in which BootSnap is trained with original training data and evaluated on denoised testing data, which is beyond the scope of the present study.

\section{Results and Discussion}\label{sec:results}
In this section, we present the experimental results; we evaluate the proposed denoiser’s impact on the SI-SDR using our synthetic testing data, as well as the performance of ridge-tracking and the integrated denoising-and-retraining pipeline for USV classification on actual field recordings.

\subsection{SI-SDR of Synthetic Testing Data}

\begin{figure}[tb]
    \centering
    % First subplot
    \begin{subfigure}[t]{0.52\linewidth}
        \centering
\includegraphics[height=0.23\textheight,width=\linewidth]{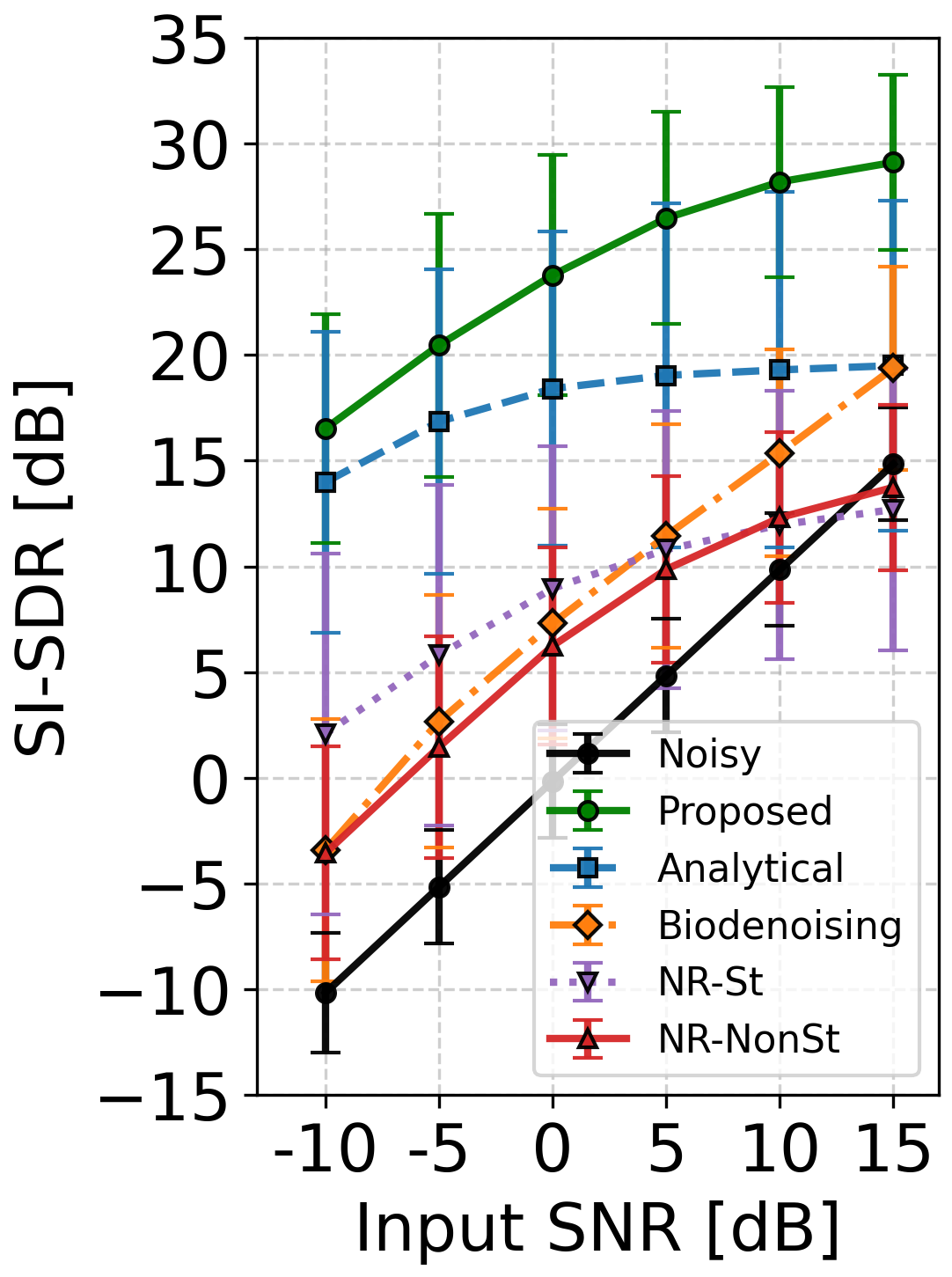}
    \label{fig:sdr_errorbars}
    \end{subfigure}
    \hfill
    % Second subplot
    \begin{subfigure}[t]{0.46\linewidth}
        \centering
\includegraphics[height=0.23\textheight,width=\linewidth]{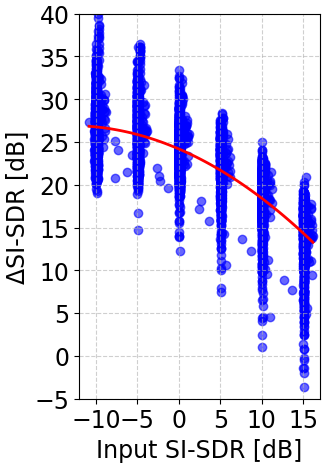}
        \label{fig:deltasdr}
    \end{subfigure}
    \caption{SI-SDR (scale-invariant signal-to-distortion ratio) before and after denoising. Left: Error bars showing the mean ± standard deviation of SI-SDR for the noisy input, our proposed denoiser, and baseline denoising methods (analytical denoiser, Biodenoising, and noisereduce (NR) under stationary (St) and non-stationary (NonSt) modes) across synthesized test samples at each input SNR level. Right: SDR improvement (\(\Delta\)SI-SDR) for individual synthesized test samples, illustrating the denoising effect of our proposed method on each signal. %Higher values indicate better SI-SDR.
    }
    \label{fig:usv_sdr}
\end{figure}

Fig. \ref{fig:usv_sdr} shows how the proposed denoiser substantially improves SI-SDR in all input SNR levels and shows SI-SDR improvement, reported over the full 200 ms signal. The mean SI-SDR increased from –10.17 dB to 16.52 dB at the lowest input SNR and from 14.84 dB to 29.11 dB at the highest input SNR. This result indicates that the proposed denoiser substantially enhances SI-SDR, with the largest gains observed at low SNR levels. 
Across all conditions, the proposed method consistently outperforms the baselines. Among the baselines, the analytical denoiser consistently outperforms Biodenoising and noisereduce. The analytical denoiser performs particularly well at low SNR of -10 dB but exhibits higher variability and a saturation effect as the input SNR increases. The noisereduce baseline performs better in the stationary than the non-stationary mode, particularly at low SNR; however, similar to the analytical denoiser, the noisereduce also saturates at higher SNRs. In contrast, Biodenoising continues to improve at all SNR levels and surpasses both noisereduce variants at higher input SNRs (10–15 dB), though it remains below the analytical denoiser. 
According to the \(\Delta\)SI-SDR plot in Fig. \ref{fig:usv_sdr}, for the proposed method, a few samples with input SNR of 15 dB show a slight reduction in SI-SDR, which is expected because these signals are at the upper edge of the network’s training range, and minor changes to already high-SNR signals can occasionally reduce SI-SDR. 

We also evaluate SI-SDR on USV intervals to assess how denoising affects the USVs themselves (results not shown). 
A similar trend is observed when evaluating SI-SDR on USV intervals. The proposed method yields the largest improvements under all conditions, increasing the mean SI-SDR of USVs from –3.50 dB to 17.01 dB at the lowest SNR and from 21.49 dB to 29.37 dB at the highest SNR. Among the baselines, the analytical denoiser achieves the highest performance, particularly at low SNR (e.g., 14.63 dB at the lowest SNR), but shows no improvement at higher SNR levels. Biodenoising exhibits more gradual improvements and reaches higher values at high SNR (e.g., from 21.49 dB to 23.79 dB), but remains below the analytical and proposed methods (e.g., from –3.50 dB to 2.8 dB). Noisereduce improves SI-SDR in both stationary and non-stationary modes, with better performance in the stationary case (e.g., 4.9 dB at the lowest SNR), but achieves lower overall SI-SDR and shows saturation at higher SNRs. 
Using the proposed denoiser, some cases of USVs with SNR above 15 dB show a slight reduction in SI-SDR after denoising. This is expected, as the proposed denoiser is not trained for such high-SNR signals, and in practice such high-SNR USVs rarely occur in actual field recordings.

\subsection{Ablation Study on USV Ridge-Tracking}

Table~\ref{tab:ridge_combined_metrics} reports evaluation metrics for fundamental ridge-tracking.
\begin{table}[t]
\caption{Evaluation metrics for fundamental ridge-tracking. Precision represents the fraction of detected ridges that correctly match manual annotations (\%), and deviation reports the mean frequency difference $\pm$ standard deviation (kHz) across all tracked points. DeepSqueak tracks ridges per \cite{coffey2019deepsqueak}, while all other methods use \textit{tfridge} in MATLAB. Bold values highlight the best performing method for each metric.}
\begin{center}
\resizebox{\columnwidth}{!}{% Resize the table to fit the page width
\setlength{\tabcolsep}{4.5pt} 
\begin{tabular}{l l l} % no vertical lines, first column left-aligned
\hline
\textbf{Method} & \textbf{Precision} & \textbf{Deviation} \\ 
\hline
Baseline (\textit{tfridge} STFT) & 94 & 2.3 $\pm$ \phantom{0}5.8 \\ 
\hline
DeepSqueak & 88.6 & 6.1 $\pm$ 10.2 \\ 
\hline
DeepSqueak Wiener RS & 92.3 & 3.5 $\pm$ \phantom{0}5.9 \\ 
\hline
\textit{tfridge} on Wiener STFT & 95 & 1.7 $\pm$ \phantom{0}4.1 \\ 
\hline
\textit{tfridge} on Wiener MTRS-STFT & 95 & 1.4 $\pm$ \phantom{0}3.4 \\ 
\hline
\makecell[l]{\textit{tfridge} on proposed denoiser \\(without loss weighting)} & 95 & 1.4 $\pm$ \phantom{0}5.1  \\
\hline
\textit{tfridge} on proposed denoiser & \textbf{96} & \textbf{0.8 $\pm$ \phantom{0}1.5} \\
\hline
\end{tabular}
}
\normalsize % Reset font size to match the main text
\end{center}
\label{tab:ridge_combined_metrics}
\end{table}
The proposed denoiser achieves the highest overall performance, with a precision of 96\% and the lowest frequency deviation of 0.8 \(\pm\) 1.5 kHz. In contrast, DeepSqueak shows the lowest performance, with a precision of 88.6\% and the largest frequency deviation of 6.1 \(\pm\) 10.2 kHz. An ablated version of the proposed denoiser, without weighted loss (i.e., the \textit{tfridge} denoiser without cRM loss weighting), maintains comparable precision but exhibits increased frequency deviation. This result indicates that loss weighting is not strictly necessary for tracking ridge points and that most points are still identified correctly, however, it also shows that loss weighting is important for accurately estimating their precise frequencies.

The results for DeepSqueak and the signal processing-based preprocessing variants (e.g., Wiener STFT and Wiener MTRS-STFT) were previously reported in~\cite{abbasi2025robust} and are included here for reference and comparison. 
%The results for DeepSqueak and signal processing-based methods (eg, wiener MTRS-STFT) are previously reported in \cite{abbasi2025robust} and are included here as a reference. 
As shown here, the MTRS-STFT method with Wiener denoising outperforms both the baseline and Wiener STFT approaches as well as DeepSqueak, achieving higher precision and also lower frequency deviation. Applying Wiener denoising and spectral reassignment to DeepSqueak (i.e., DeepSqueak Wiener RS) improves its precision and frequency accuracy.

\begin{figure}[tb]
    \centering
\includegraphics[width=\columnwidth]{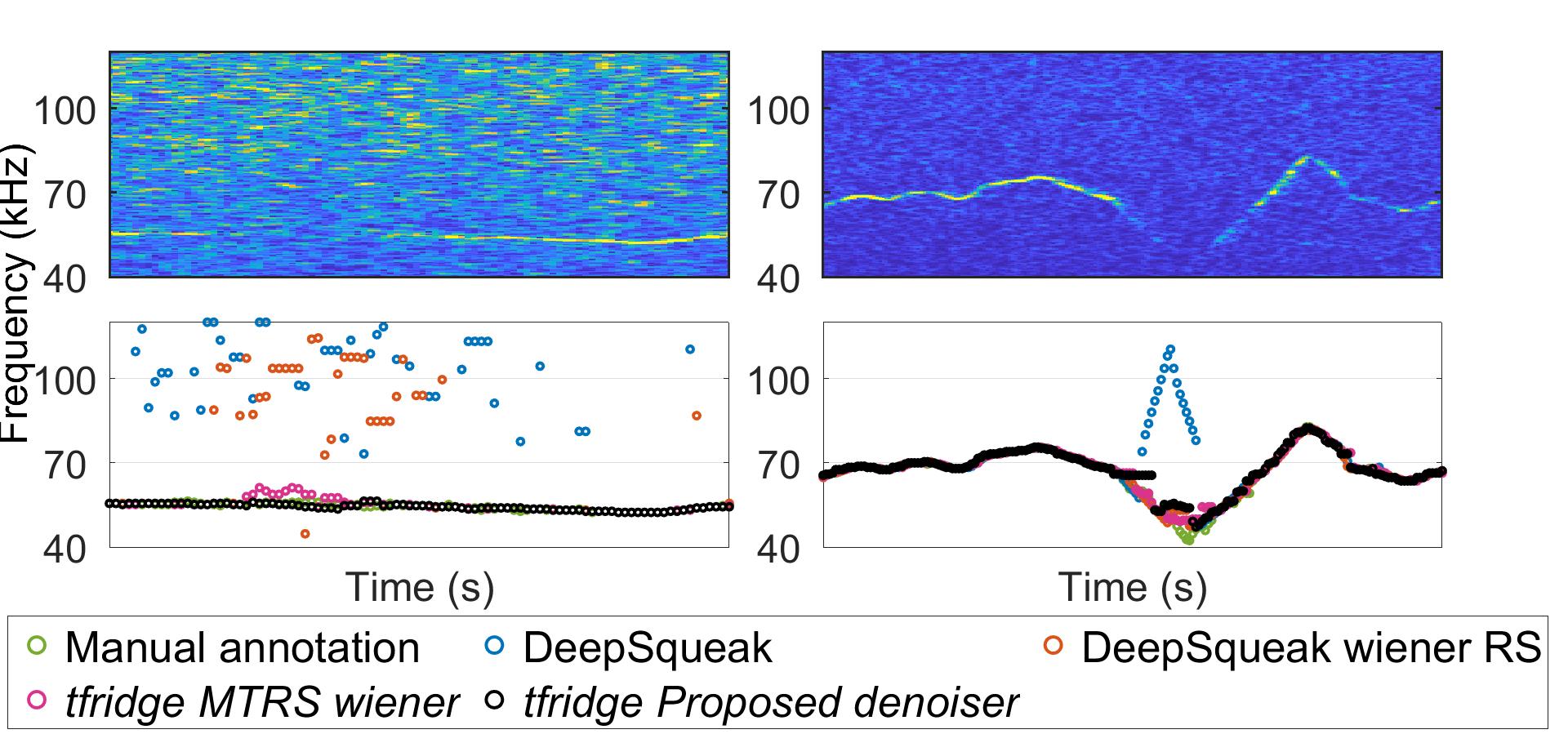}

    \caption{Spectrograms of two example USVs (top) with corresponding fundamental ridge-tracking results (bottom) relative to manual annotations. Time-frequency representations are computed using the STFT. Ridge tracking for DeepSqueak-based methods follows the DeepSqueak approach, whereas for the other methods, tracking is performed with MATLAB’s \texttt{tfridge} function applied to signals pre-processed with different denoising or enhancement approaches.}\label{fig:fundamental_ridge_comparison}
\end{figure}

Fig. \ref{fig:fundamental_ridge_comparison} presents the results of ridge-tracking using different methods. The first example shows a USV where the proposed method outperforms the MTRS Wiener method. The second example shows a USV with a faint segment in the middle, where all methods struggle. Nonetheless, the proposed denoising method yields the lowest error in ridge-tracking compared to the other methods.
%%%%%%%%%%%%%%%%
% fundamental_ridge_comparison
%%%%%%%%%%%%%%%%%
%%%%%%%%%%%%%%%%
% harmonic component tracking
%%%%%%%%%%%%%%%%%

\begin{table}[tb]
\caption{Evaluation metrics for the tracking of harmonic partials. Precision and recall are reported as percentages (\%), and deviation indicates the mean frequency difference $\pm$ standard deviation (kHz) for all tracked points. Harmonic partial tracking is performed for the input representations listed in the table using the autocorrelation-based method described in Section \ref{sec:ridge_tracking_init}. Bold values indicate the best-performing method for each metric.}
\begin{center}
\resizebox{\columnwidth}{!}{% 
\setlength{\tabcolsep}{4.5pt}  % Reduce column space
\begin{tabular}{l l l l} % all columns left-aligned
\hline
\textbf{Input} & \textbf{Precision} & \textbf{Recall} & \textbf{Deviation} \\
\hline
Baseline (CQT) & 73 & 51 & 1.0 $\pm$ 0.3 \\
\hline
Wiener CQT & 81 & 54 & 0.8 $\pm$ 0.3 \\
\hline
Wiener MTRS-CQT& 89 & 58 & 0.8 $\pm$ 0.1 \\
\hline
\makecell[l]{proposed denoiser \\(without loss weighting)} & \textbf{97} & 66 & 0.8 $\pm$ 0.3 \\
\hline
proposed denoiser & 93 & \textbf{83} & 0.9 $\pm$ 0.2 \\
\hline
\end{tabular}
}
\normalsize % Reset font size to match the main text
\end{center}
\label{tab:harm_ridge_combined_metrics}
\end{table}
The results of the harmonic partial tracking are summarized in Table~\ref{tab:harm_ridge_combined_metrics}. Applying the proposed denoiser to the audio signal achieves the best overall performance, with the highest recall (83\%) and precision (93\%), while maintaining low frequency deviation (0.9 \(\pm\) 0.2 kHz). In contrast, the baseline representation (i.e., CQT) without denoising shows the lowest performance, with a precision of 73\%, a recall of 51\%, and a deviation of 1.0 \(\pm\) 0.3 kHz. Wiener denoising improves the precision to 81\% and slightly reduces the deviation to 0.8 \(\pm\) 0.3 kHz, but recall remains limited. Using the multitaper reassigned CQT with Wiener denoising (MTRS-CQT) increases the precision to 89\% and keeps the deviation low (0.8 \(\pm\) 0.1 kHz), with only modest improvement in the recall.

The proposed denoiser without cRM loss weighting achieves the highest precision (97\%) but lower recall (66\%), indicating occasional missed low-amplitude harmonic partials. The proposed method balances precision and recall (93\% and 83\%, respectively), successfully tracking weaker harmonic partials without increasing frequency deviation or false positives.
%%%%%%%%%%%%%%%%
%Denoising Impact on the F1-score of the USV Classifier
%%%%%%%%%%%%%%%%%%%%%%%%
%\subsection{Denoising Impact on the F1-score of the USV Classifier}
\subsection{USV Classification Performance with Denoised Training Data}
\begin{figure*}[t]  % Use figure* for full-width in 2-column layout
  \centering\includegraphics[width=0.85\textwidth]{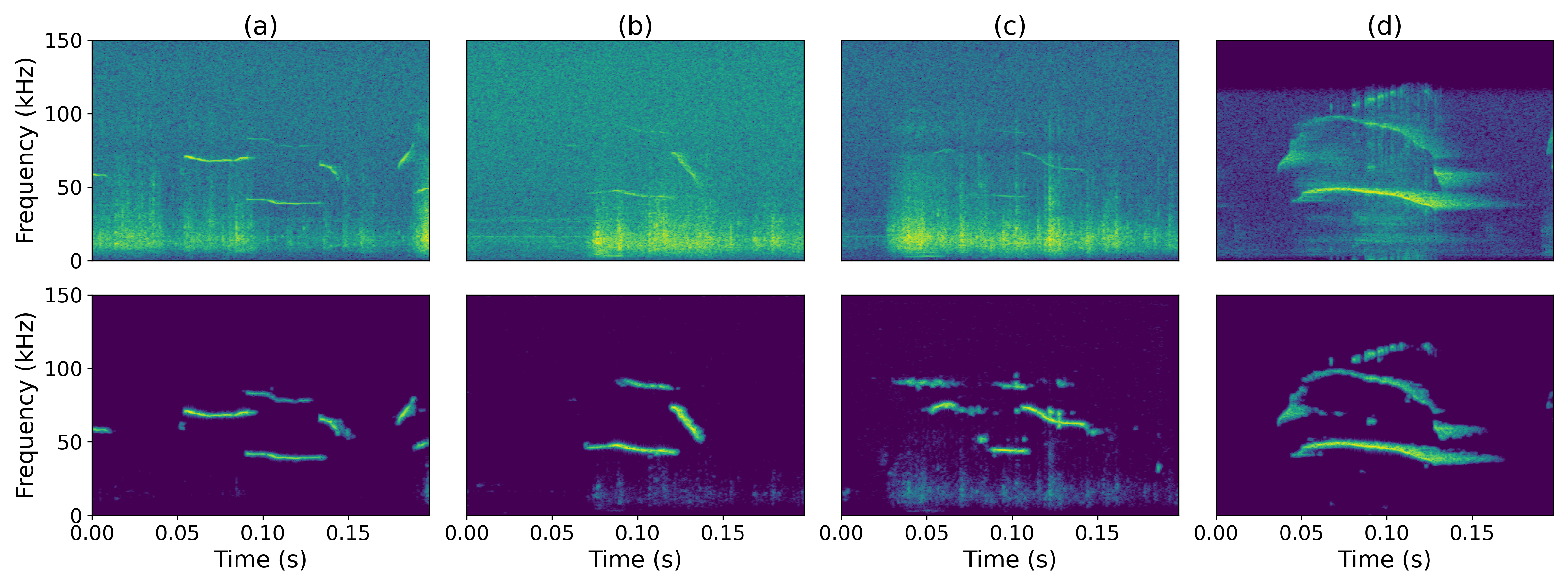}
  \caption{Spectrograms of the original (top) and the denoised (bottom) audio from four datasets: (a) training (wild mouse, correct ridge), (b) testing (wild mouse, correct ridge), (c) testing (wild mouse, incorrect ridge), and (d) testing (domesticated mouse), all from syllable class C3. These spectrograms show samples from the datasets used for training and evaluating the BootSnap classifier. Time is shown in seconds and frequency in kHz. Denoising is performed using the proposed method. Color intensity represents signal amplitude.}\label{fig:original_denoised_grid}
\end{figure*}

The first three columns of Fig.~\ref{fig:original_denoised_grid} show spectrograms of examples of one USV syllable type (C3) from wild mice. The spectrograms elucidate examples from the training (Fig.~\ref{fig:original_denoised_grid} (a)) and testing (Fig.~\ref{fig:original_denoised_grid} (b),(c)) datasets, including the correct-ridge (Fig.~\ref{fig:original_denoised_grid} (a),(b)) and the incorrect-ridge (Fig.~\ref{fig:original_denoised_grid} (c)) data (see Table~\ref{tab:data_dist}). Samples from the correct-ridge subset have a higher SNR than those from the incorrect-ridge subset, which shows that the proposed denoiser yields cleaner outputs on the correct-ridge data. However, when segments are very faint, the proposed denoiser can barely preserve them; and although they are present in the denoised output, they remain extremely weak (Fig.~\ref{fig:original_denoised_grid} (b)). The last column (Fig.~\ref{fig:original_denoised_grid} (d)) shows a USV spectrogram from a domesticated strain, which may show noise patterns different from those of wild mice, as these recordings are made in different laboratories, originate from very different mice, and have more pronounced echo effects. Despite these novel conditions, the proposed denoiser effectively reduces noise.

%%%%%%%%%%%
% denoiser output
%%%%%%%%%%%%%%  

Table~\ref{tab:bootsnap_wild_results} reports the USV classification performance for the three BootSnap models described in Section~\ref{sec:eval_classification}, evaluated with the correct-ridge and incorrect-ridge testing subsets. 

%%%%%%%%%%%%%%%%5
% classification results wild mice
%%%%%%%%%%%%%%%%%%%%
\begin{table*}[ht]
\centering
\normalsize
\caption{USV classification performance for three BootSnap models trained on original, analytically denoised, and proposed-denoised audio. All training conditions additionally use BootSnap’s built‑in default denoising to maintain pipeline compatibility. Models are evaluated on the correct-ridge and incorrect-ridge wild-mice testing subsets. Performance is reported as class-wise F1 and macro-F1 scores (\%). Bold values indicate the best‑performing model.}
\label{tab:bootsnap_wild_results}
\begin{tabular}{@{}l l rrrrrrrr r@{}}
\hline
\textbf{Input Condition} & \textbf{Dataset Type} & \multicolumn{8}{c}{\textbf{Class-wise F1-Score (\%)}} & \textbf{macro-F1 (\%)} \\
\cline{3-10}
 & & C & C2 & C3 & D & F & U & UI & UP & \\
\hline
Original & correct-ridge wild mice & 85 & 49 & 92 & 84 & 88 & 85 & 89 & 93 & 83 \\
Denoised (analytical denoiser) & correct-ridge wild mice & 84 & 58 & 89 & 83 & 87 & 81 & 88 & 91 & 83 \\
Denoised (proposed denoiser) & correct-ridge wild mice & \textbf{90} & \textbf{77} & \textbf{95} & \textbf{89} & \textbf{89} & \textbf{88} & \textbf{92} & 93 & \textbf{89} \\
\hline
Original & incorrect-ridge wild mice & \textbf{76} & 33 & 81 & 70 & 77 & 64 & 75 & 79 & 69 \\
Denoised (analytical denoiser) & incorrect-ridge wild mice & 70 & 25 & 65 & 60 & 71 & 61 & 63 & 68 & 60 \\
Denoised (proposed denoiser) & incorrect-ridge wild mice & 74 & \textbf{40} & \textbf{82} & \textbf{71} & \textbf{79} & \textbf{71} & 75 & \textbf{82} & \textbf{72} \\
\hline
\end{tabular}
\end{table*}
In the correct-ridge testing subset, the classifier trained on data processed with the proposed denoiser achieves a higher macro-F1 (89\%) compared to 83\% for both the original and analytically denoised conditions. While training classifier on analytically denoised data improves performance for some classes (e.g., C2), it does not yield an overall macro-F1 gain. 
In contrast, training on data processed with the proposed denoiser improves performance across most classes, with the most notable improvement observed for class C2. This is likely because C2 represents USVs that often have very small jump segments and are frequently buried in noise; denoising makes these subtle jumps more detectable, allowing the classifier to perform better. 

For the incorrect-ridge subset, training on analytically denoised data reduces the macro-F1 from 69\% (original) to 60\%, indicating that errors in ridge estimation can introduce distortions that negatively affect classification under low-SNR conditions. In contrast, training the classifier on data processed with the proposed denoising approach increases the macro-F1 score to 72\%, indicating improved robustness under such challenging conditions. %demonstrating that denoising also enhances the classifier’s generalizability. 
Although the performance on the incorrect-ridge data remains lower, training on proposed-denoised data mitigates this drop — particularly for classes U and C2 — highlighting increased robustness to domain shift.%its contribution to the model robustness under a domain shift. 

Table~\ref{tab:bootsnap_lab_results} summarizes the performance of the three BootSnap classifiers on the domesticated strain dataset.
%%%%%%%%%%%%%%%%5
% classification results lab mice
%%%%%%%%%%%%%%%%%%%%
\begin{table*}[t]
\centering
\normalsize
\caption{USV classification performance of the three BootSnap models %trained with and without the proposed denoiser
 trained on original, analytically denoised, and proposed-denoised audio. All training conditions additionally use BootSnap’s built‑in default denoising to maintain pipeline compatibility. Models are evaluated on a domesticated-mouse dataset. The performance is reported as class-wise and macro-F1 scores (\%). Bold numbers indicate the model with better performance.}
\label{tab:bootsnap_lab_results}
\begin{tabular}{@{}l l rrrrr r@{}}
\hline
\textbf{Input Condition} & \textbf{Dataset Type} & \multicolumn{5}{c}{\textbf{Class-wise F1-Score (\%)}} & \textbf{macro-F1 (\%)} \\
\cline{3-7}
 & & C & C2 & C3 & UI & Rise & \\
\hline
Original &Domesticated mice& 34 & 36 & \textbf{80} & 51 & 83 & 57 \\
Denoised (analytical denoiser)&Domesticated mice& \textbf{39} & 49 & 75 & 55 & 80 & 60 \\
Denoised (proposed denoiser)& Domesticated mice& 28 & \textbf{50} & 78 & \textbf{62} & \textbf{85} & \textbf{61} \\
\hline
\end{tabular}
\end{table*}
% \subsection{Feature Preservation Evaluation}
Training the classifier on data processed with the proposed denoiser (the proposed integrated pipeline) achieves the highest macro-F1 (61\%), compared with 57\% for the original input and 60\% for training on analytically denoised data, indicating improved robustness on data unseen by both the denoiser and the classifier.
The improvement is again most pronounced for class C2 (36\% to 50\%) but this time also for UI (51\% to 62\%), while the class Rise also shows a smaller gain (an increase of 2\%). 
Training on analytically denoised data also improves overall performance (to 60\%) and yields gains for several classes, although these improvements are generally smaller than those obtained with the proposed integrated pipeline. In particular, class C shows an improvement with the analytical denoiser (34\% to 39\%), suggesting its patterns are sufficiently preserved under analytical denoising.

However, with the proposed integrated pipeline, class C exhibits a decrease (34\% to 28\%),= and class C3 shows a slight reduction (80\% to 78\%); however, the latter remains high, suggesting a limited sensitivity to denoising. The reduction for syllable type C might be attributed to subtle alterations in its spectral features caused by denoising, which can slightly affect the classifier’s ability to recognize this class. 

\subsection{Discussion}
Our results show that the proposed denoising method is effective in multiple tasks, including SI-SDR on synthetic testing data, ridge-tracking, and improved classification performance when retraining on denoised data. The results also demonstrate consistent improvements over baseline denoising approaches.
Below, we discuss the significance of these findings, the practical scope and limitations of the method, and directions for future work.

%%%%%%%%%%%%
%sisdr
%%%%%%%%%%%%%%%

The proposed denoiser substantially improves SI-SDR, with the largest gains observed at low-SNR levels, which are the most challenging in practice. Compared to baseline denoisers, it consistently outperforms all alternatives across SNR levels.%Moreover, the denoised output can preserve multiple ridge trajectories. This is particularly relevant in recordings containing overlapping vocalizations and may facilitate future source-separation approaches \cite{denton2022improving}.
These improvements indicate that previously faint or noisy USVs can be more accurately recovered, supporting downstream tasks such as ridge-tracking and classification. %As shown in Fig. \ref{fig:original_denoised_grid}, some very faint signals buried in noise are still difficult to recover, which could be addressed in future work. 
Beyond objective metrics like SI‑SDR, perceptual evaluation of denoised bioacoustic signals remains challenging, particularly for synthesized vocalizations for which no perceptual ground truth exists. 
%More broadly, preservation of multiple ridge trajectories in the denoised outputs may support future studies of bioacoustic signal analysis, including overlapping vocalizations and bioacoustic source separation \cite{denton2022improving}.

Among the baselines, the analytical denoiser performs best, suggesting that ridge-guided signal modeling captures key characteristics of USVs, although its performance saturates at higher SNR and it is sensitive to ridge-estimation errors under high-noise conditions. Specifically, it is affected by erroneous ridge jumps, weak signal components, mask choice, overlapping noise, and phase-retrieval distortions, all of which limit its effectiveness. Biodenoising shows a more gradual improvement across the SNR levels but remains below the analytical denoiser. This may reflect domain mismatch, as the pretrained model is optimized for speech signals and is applied here without USV-specific fine-tuning. While retraining on USV-adapted pseudo-clean targets could potentially improve performance, such targets may still be suboptimal for preserving the narrowband tonal nature of USVs. In addition, the speech-enhancement model used in Biodenoising may further limit its ability to capture USV-specific structures. Noisereduce achieves its largest gains at low SNR, particularly in the stationary mode, but also saturates at higher SNR, reflecting limitations of noise-estimation-based methods.

%%%%%%%%%%%%
%fundamental ridge tracking
%%%%%%%%%%%%%%%
Compared with existing methods for USV analysis, the proposed denoising approach results in substantially lower frequency deviation in fundamental ridge-tracking (0.8\(\pm\)1.5 kHz vs. 1.4\(\pm\)3.4 kHz), while maintaining a comparable precision (96\% vs. 95\%).
%%%%%%%%%%
% classical vs U-Net method
%%%%%%%%%
Moreover, although classical signal enhancement methods — such as multitaper reassigned CQT combined with Wiener filtering — can improve spectral clarity, the proposed approach demonstrates a superior performance in harmonic partial ridge-tracking, achieving higher precision (93\% vs. 89\%) and recall (83\% vs. 58\%).

Our ablation study highlights the critical role of the ridge-guided loss in the U-Net. Removing the ridge‑guided weights leads to over-suppression of faint fundamental and harmonic partial components, degrading ridge-tracking performance. Incorporating these weights enhances performance by guiding the denoising network to focus on ridges, thereby improving the detection of faint components.

However, the current evaluation of ridge-tracking is based on a relatively small dataset of 193 USVs. Increasing the number of manually annotated samples would enable a more comprehensive and representative assessment. Manual ridge annotations might also be affected by noise and weak signal content. To address this, synthetic data generated from known ridges could offer an objective, complementary benchmark for evaluating the accuracy of the ridge-tracking.

%%%%%%%%%%%%%%%
%wild mice result:
%%%%%%%%%%%%%%
We assess the impact of the integrated denoising-and-retraining pipeline on USV classification using two subsets from wild house mice and one subset from a domesticated strain of mice. The wild mouse data include a subset with noise levels similar to the training data (the “correct-ridge” subset), which serves as an in-sample testing data, and another subset with substantially higher noise (the “incorrect-ridge” subset), which serves as an out-of-sample testing data. The domesticated strains mouse data also serves as another out-of-sample test condition, as it is not used during training for either the proposed denoiser or the classifier.

In the wild mouse data, although the proposed integrated pipeline improves the F1 scores in both subsets, the gains are smaller for the noisier data (6\% vs. 3\%). This reduced improvement likely results from several factors, including the challenges posed by extremely low-amplitude USVs and inconsistencies in manual annotations. In some recordings, segments of USVs are nearly indistinguishable due to their extremely low amplitude; in others, inconsistencies in manual annotations may limit the achievable classification accuracy. Manual USV annotation is inherently subjective, and prior work~\cite{abbasi2022capturing} has reported inter-observer reliability of 80–84\% for an eight-class classification of wild mouse USVs, suggesting a practical upper bound on classifier performance. This issue could be reduced by using our proposed method to provide enhanced visual cues during manual annotation, as mentioned above. 
By comparison, training the classifier on analytically denoised data yields less consistent improvements, particularly under high-noise conditions.
%In contrast, the U-Net leverages global signal context and learns a robust mapping from noisy inputs to clean outputs, allowing it to smooth over artifacts and adapt to variability in the data, making it more effective under challenging conditions.
This is largely due to the analytical denoiser’s sensitivity to ridge-estimation errors and its limited ability to handle variability in noisy data. In contrast, the U-Net leverages global signal context and learns a robust mapping from noisy inputs to clean outputs, allowing it to smooth over artifacts and adapt to variability in the data, making it more effective under challenging conditions.
%%%%%%%%%%%%%%%
%lab mice result:
%%%%%%%%%%%%%%

The domesticated mouse dataset represents a realistic case of domain shift. Results show an increase in macro-F1 from 57\% to 61\%, demonstrating that proposed integrated pipeline improves the classifier generalization to even these markedly different mice. These findings underscore the value of signal enhancement for improving the robustness under domain shift, particularly when retraining the classifier on out-of-sample data is impractical.

\section{Conclusion}\label{sec:discussion}
Bioacoustic signals, such as bat echolocation calls, dolphin whistles, bird songs, and mouse ultrasonic vocalizations, are often degraded by ambient noise. While advanced denoising models can suppress complex noise, they typically require clean ground truth data for supervised training, which is rarely available in bioacoustics. To address this limitation, we first used ridge information for the synthesis of a training dataset. We then trained a U-Net denoiser to predict a complex ratio mask using a ridge-guided loss. 

Using mouse USV recordings as a case study, we showed that the proposed method substantially improves SI-SDR of the synthesized test set (e.g., from –10.17 dB to 16.52 dB at the lowest input SNR). It also enhanced ridge-tracking accuracy on actual field recordings (e.g., reducing the frequency deviation of the fundamental ridge from 1.4 $\pm0.5$ 3.5 kHz using \cite{abbasi2025robust} to 0.8 $\pm0.5$ 1.5 kHz). Furthermore, the integrated denoising‑and‑retraining pipeline increased USV classification performance on actual field recordings under both in-sample (e.g., macro-F1 from 83\% with the original recordings to 89\%) and out-of-sample (macro-F1 from 69\% to 72\%) noise conditions.

Many animal vocalizations exhibit ridges in their time–frequency representations, and there is growing interest in extracting ridges from these bioacoustic signals \cite{best2025bioacoustic}.  Although we evaluated proposed method using mouse USVs as a case study, it is broadly applicable to other bioacoustic signals containing ridge-like structure, such as tonal and locally sinusoidal vocalizations. Moreover, the preservation of multiple ridge trajectories in denoised outputs suggests potential applicability of the proposed method to source separation tasks, such as disentangling overlapping signals in complex acoustic scenes, which are common in natural environments \cite{denton2022improving}. The method is not tied to a specific ridge-tracking algorithm; alternative ridge-tracking approaches—including fundamental frequency estimators benchmarked in \cite{best2025bioacoustic}—can be incorporated to better account for the characteristics of different species and signal types.

Despite its strengths, the method has limitations. Very faint calls buried in noise may remain weak after denoising, as shown in Fig. \ref{fig:original_denoised_grid}, and very high-SNR synthetic signals can occasionally lose SI-SDR after denoising. Additionally, analytical reconstruction is sensitive to ridge-estimation errors under high-noise conditions, which can introduce artifacts into the synthesized audio. These challenges highlight conditions under which the method struggles to recover faint signals. Furthermore, the proposed denoiser is inherently limited to signals with well-defined ridge structures; broadband or strongly noise-like vocalizations, where ridges are not well defined, fall outside the scope of the present work.

Within this scope—signals with identifiable ridges—the approach supports more reliable feature extraction and downstream analysis across taxa, with possible implications for playback experiment, automated species identification, behavioral analysis, and conservation monitoring.

\section*{Acknowledgment}
This work was supported by the FWF project P36446-B (S. Zala, R. Abbasi, D. Penn, and P. Balazs), FWF project P34922-N and the WWTF project LS23024 (P. Balazs and C. Hollomey), P28141-B25 (D. Penn and S. Zala), the WWTF project LS23-014 (N. Holighaus), and the ANR project ANR-23-CE37-0025 (V. Lostanlen). The computational results presented are obtained using the CLIP cluster \footnote{\url{https://www.clip.science}}.
%The authors are ordered according to their contribution.
\ifCLASSOPTIONcaptionsoff
  \newpage
\fi


%\bibliographystyle{IEEEtran}


% \end{thebibliography}
\bibliographystyle{IEEEtran}  % IEEEtran style
\bibliography{sample,biblioall}


% \begin{IEEEbiographynophoto}{Michael Shell}


%\newpage



%\vfill


\end{document}